\documentclass[prd,onecolumn,showpacs,floatfix,superscriptaddress,nofootinbib]{revtex4}
\usepackage{graphicx}
\usepackage{epsfig}
\usepackage{bm}
\usepackage{amssymb}
\usepackage{float}
\usepackage{amsmath}
\usepackage{dcolumn}
\usepackage{cancel}
\usepackage[colorlinks]{hyperref}
\usepackage[usenames,dvipsnames]{color}
\hypersetup{
     breaklinks=true,
    pdfstartview={FitH},    
    colorlinks=true,       
    linkcolor=blue,          
    citecolor=red,        
    filecolor=magenta,      
    urlcolor=blue,           
    anchorcolor=green,      
    linktocpage=true
}

\def\doi{http://doi.org}

\def\be{\begin{equation*}}
\def\ee{\end{equation*}}

\begin{document}

\title{Quasinormal modes and their anomalous behavior for  black holes in $f(R)$ gravity}


\author{Almendra Arag\'{o}n}
\email{almendra.aragon@mail.udp.cl} \affiliation{Facultad de
Ingenier\'{i}a y Ciencias, Universidad Diego Portales, Avenida Ej\'{e}rcito
Libertador 441, Casilla 298-V, Santiago, Chile.}

\author{P. A. Gonz\'{a}lez}
\email{pablo.gonzalez@udp.cl} \affiliation{Facultad de
Ingenier\'{i}a y Ciencias, Universidad Diego Portales, Avenida Ej\'{e}rcito
Libertador 441, Casilla 298-V, Santiago, Chile.}

\author{Eleftherios Papantonopoulos}
\email{lpapa@central.ntua.gr} \affiliation{Physics Division,
National Technical University of Athens, 15780 Zografou Campus,
Athens, Greece.}

\author{Yerko V\'asquez}
\email{yvasquez@userena.cl}
\affiliation{Departamento de F\'isica y Astronom\'ia, Facultad de Ciencias, Universidad de La Serena,\\
Avenida Cisternas 1200, La Serena, Chile.}

\date{\today }

\begin{abstract}

We study the propagation of scalar fields in the background of an asymptotically de-Sitter black hole solution in $f(R)$ gravity. The aim of this work is to analyze  in 
modified theories  of gravity the existence of an anomalous decay rate of the quasinormal modes (QNMs) of a massive scalar field which was recently reported in Schwarzschild black holes backgrounds, in which the longest-lived modes are the ones with higher angular number,
for a scalar field mass smaller than
a critical value, while that beyond this value the
behavior is inverted.
We study the QNMs for various overtone numbers
and they depend on a parameter $\beta$ which appears in the metric and characterizes the  $f(R)$ gravity. For small $\beta$, i.e small deviations from the  Schwarzschild-dS black hole the anomalous behaviour in the QNMs is present for the photon sphere modes, and the critical value of the mass of the scalar field depends on the parameter $\beta$ while  for large $\beta$, i.e large deviations the anomalous behaviour and the critical mass does not appear. Also, the critical mass of the scalar field  increases when the overtone number increases until the $f(R)$ gravity parameter $\beta$ approaches the  near extremal limit at which  the critical mass of the scalar field does not depend anymore on the overtone number. The imaginary part of the quasinormal frequencies is always negative leading to a stable  propagation of the scalar fields in this background.

\end{abstract}


\maketitle
\flushbottom

\tableofcontents

\newpage

\section{Introduction}

Modified theories of gravity in which  the Einstein-Hilbert action is replaced with a generic form of $f(R)$ gravity have been introduced in a attempt to  describe the early and late cosmological evolution.  Another motivation to study such theories is the understanding of the existence of  dark energy and dark matter consistent with the recent  observations,  without introducing  new material ingredients that have not yet been detected by experiments \cite{Nojiri:2006ri,Copeland:2006wr,Nojiri:2010wj,Clifton:2011jh,Capozziello:2007ec}. Modifying the action not only affects the dynamics of the universe, it can also alter the dynamics at the galactic or solar system scales. Therefore, modified theories of gravity with  curvature corrections, provide a deeper understanding of General Relativity (GR).

Assuming that the  gravitational Lagrangian is not only  a linear function of $R$, variable modified theories of gravity were considered describing the cosmic evolution at early times, in which the  gravitational Lagrangian contains some of the four possible second-order curvature invariants. Also models in which higher order curvature invariants as functions of the Ricci scalar were introduced in the  gravitational Lagrangian  resulting to various $f(R)$ gravity models \cite{DeFelice:2010aj}-\cite{Vainio:2016qas}.  Although such theories exclude contributions from any curvature invariants other than $R$,  they could also avoid the Ostrogradski instability \cite{Ostrogradsky:1850fid} which proves to be problematic for general higher derivative theories \cite{Woodard:2006nt}.

One of the first modifications of the Einstein Lagrangian density was proposed in \cite{Buchdahl}. A more natural modification of the Einstein Lagrangian is to add terms $R^n$,   like the Starobinsky model $f(R)=R+\alpha R^2$ \cite{Starobinsky:1980te}. For $n<0$ such corrections become important in the late universe and can lead to self-accelerating vacuum solutions \cite{Carroll:2003st,Carroll:2003wy,Capozziello:2002rd,Capozziello:2003gx}. However, these models suffer from instabilities \cite{Soussa:2003re,Faraoni:2005vk} and there are strong constraints from the solar system \cite{Chiba:2003ir}.  A wide range of phenomena can be explained by considering different $f(R)$ functions. Some discussions have been performed, such as on gravitational wave detection  \cite{Corda:2008si,Corda:2010zza},  early-time inflation \cite{Bamba:2008ja}, cosmological phases \cite{Nojiri:2006gh,Nojiri:2006be,Nojiri:2008fk}, the singularity problem \cite{Kobayashi:2008wc}, the stability of the solutions
\cite{Faraoni:2005ie,Capozziello:2006dj,Amendola:2006we}, and other different branches have been studied
\cite{Akbar:2006mq}.

Spherically symmetric static solutions have been studied in $f(R)$ theories of gravity.
In particular, it was shown that for a large class of models the Schwarzschild-de Sitter (SdS) metric is an exact solution of the field equations. However, in addition to the SdS metric, $f(R)$ theories typically also have new different solutions \cite{Multamaki:2006zb}.
Also, static spherically symmetric perfect fluid solutions have been studied by matching the outside SdS-metric to the metric inside the mass distribution that leads to additional constraints that limit the allowed fluid configurations \cite{Multamaki:2006ym}, while in  \cite{delaCruzDombriz:2009et} black hole solution were found  with and without electric charge, and  exact spherically symmetric solutions were discussed in \cite{Hendi:2011eg,Sebastiani:2010kv,Hendi:2012nj,Asgari:2011ix,Ghosh:2012ji,Hendi:2014mba}. Also, exact charged black hole solutions in $f(R)$ gravity theories with dynamic curvature in $D$-dimensions were discussed in \cite{Tang:2019qiy}.

A way to probe the behaviour and the stability of black holes is the study of their quasinormal modes (QNMs) and quasinormal frequencies (QNFs)
\cite{Regge:1957td,Zerilli:1971wd, Kokkotas:1999bd, Nollert:1999ji, Konoplya:2011qq}. The QNMs depend on the black hole parameters and probe field parameters, and on the fundamental constants of the system and they are independent of the initial conditions of the perturbations. The QNMs are described by complex frequencies, $\omega=\omega_R+i \omega_I$, in which the real part $\omega_R$ determines the oscillation timescale of the modes, while the complex part $\omega_I$ determines their exponential decaying timescale (for a review on QNM modes see \cite{Kokkotas:1999bd, Berti:2009kk}).

 The QNMs and QNFs have been calculated using various numerical and analytical techniques \cite{Finazzo:2016psx, Gonzalez:2017shu, Abdalla:2019irr, Gonzalez:2018xrq, Becar:2019hwk, Aragon:2020qdc}. In the case of gravitational perturbations
 it was found that for the Schwarzschild and Kerr black hole background the longest-lived modes are always the ones with lower angular number $\ell$.  In the case of a massive probe scalar field it was found \cite{Konoplya:2004wg, Konoplya:2006br,Dolan:2007mj, Tattersall:2018nve},  at least for the overtone $n = 0$, that if the mass of the  scalar field is small, then the longest-lived QNMs are those with a high angular number $\ell$, whereas if the mass of the scalar scalar is large the longest-lived modes are those with a low angular number $\ell$. This inverted behaviour appears when  the mass of the scalar field exceeds a critical value,  which corresponds to the value of the scalar field mass where the behavior of the decay rate of the QNMs is inverted and can be obtained from the condition $Im(\omega)_{\ell}=Im(\omega)_{\ell+1}$ in the {\it eikonal} limit, that is when $\ell \rightarrow \infty$.
This anomalous decay rate for small mass scale of the scalar field was recently discussed in \cite{Lagos:2020oek}.

The study of the anomalous behaviour of QNMs was extented to other asymptotic geometries, such as, Schwarzschild-de Sitter and Schwarzschild-AdS black holes in \cite{Aragon:2020tvq}. It was found that the same behaviour is present in  the Schwarzschild-de Sitter background, i.e, the absolute values of the imaginary part of the QNFs decay when the angular harmonic numbers increase if the mass of the scalar field is smaller than the critical mass, and they grow when the angular harmonic numbers increase, if the mass of the scalar field is larger than the critical mass. Also it was found that the increase of the  value of the cosmological constant results in the increase of the value of the critical mass of the scalar field. It was also found that the anomalous behaviour is present in Schwarzschild-AdS black holes backgrounds; however, there is not a critical mass where the behavior in the decay rate is inverted. Also anomalous decay of QNMs in accelerating black holes was recently reported in \cite{Destounis:2020yav}, as well as, for Reissner-Nordstr\"om black holes \cite{Fontana:2020syy}.  Furthermore, the anomalous behaviour of QNMs was observed for fermionic field perturbations in Schwarzschild-de Sitter black holes \cite{Aragon:2020teq}.

Most of the $f(R)$ models have a very interesting property. If a conformal transformation is carried out from the original  Jordan frame to the Einstein frame then,  the conformal metric appears to be coupled minimally to gravity and a new scalar field appears which is  coupled minimally to the conformal metric and also a scalar potential is generated. The resulted theory can be considered as a scalar-tensor theory with  a geometric (gravitational) scalar field. Then it was shown in \cite{Canate:2015dda,Canate:2017bao}, that this geometric scalar field cannot dress a $f(R)$ black hole with  hair, therefore, the non-hair theorem is respected in these models. Matter was introduced in \cite{Sultana:2018fkw} and a
$R^2$ gravity model was discussed having a conformally coupled scalar field with associated Higgs-like potential and  a static spherically symmetric black hole was found. However, the conformally coupled scalar field could not give to the black hole hair, because it is regulated by the same parameters that appear in the metric. A scalar field minimally coupled to gravity with a self-interacting potential in a $f(R)$ theory was discussed in \cite{Karakasis:2021lnq}. An exact hairy black hole solution was found similar to the BTZ black hole. Also  the process of scalarization of $f(R)$ gravity theories was investigated in \cite{Tang:2020sjs}.

As we already discussed the knowledge of the QNMs and QNFs can give us vital information of the behaviour, properties and the stability of black holes. In this work we will consider  a specific  $f(R)$ theory which accepts a black hole solution  \cite{Saffari:2007zt}. In this theory the Ricci scalar receives non-linear correction terms which they introduce new parameters in the metric functions of the  black hole solution, expressing the change   of the curvature altering the strength of  the gravitational force. In this theory we will consider a massless and a  massive test scalar field scattered off the background black hole and we will calculate the QNMs and QNFs. The motivation of this work is to calculate the scalar field perturbations and study the behaviour of QNMs and QNFs and see how they are effected by the presence of non-linear curvature terms which appear explicitly in the metric functions. 

The manuscript is organized as follows: In Sec. \ref{background} we give a brief review of the model considered for $f(R)$ gravity and its black hole solution. In Sec. \ref{QNM}, we study the scalar field stability and we calculate the QNFs of scalar perturbations numerically by using the pseudospectral Chebyshev method for each family of modes.
In Sec. \ref{WKBJ},  we perform an analysis using the WKB method to get some analytical insight.
Finally, our conclusions are in Sec. \ref{conclusion}.

\section{$f(R)$ Modified Gravity}
\label{background}
We consider a  generic action in $f(R)$ gravity depending on the Ricci scalar $R$  given by
\begin{equation}
    S=\frac{1}{2k}\int d^4x\sqrt{-g}f(R)+S_m~,
\end{equation}
where $S_m$ is the matter content of the theory. In \cite{Saffari:2007zt} a specific ansatz for the function $f(R)$ was considered
\begin{equation}
    f(R)=R+\Lambda+\frac{R+\Lambda}{R/R_0+2/\alpha}\ln{\frac{R+\Lambda}{R_c}}\,,
    \label{fRfunct}
\end{equation}
where $\Lambda$ corresponds to the cosmological constant, $R_c$ is a costant of integration and $R_0=6\alpha^2/d^2$, with $\alpha$ and $d$ being free parameters. Then considering a  spherically symmetric metric
\begin{equation}
    ds^2=B(r)dt^2-A(r)dr^2-r^2(d\theta^2+sin^2\theta d\phi^2)\,,
    \label{metric}
\end{equation}
where $A(r)=B(r)^{-1},$  a vacuum spherically symmetric solution was found in \cite{Saffari:2007zt} with $B(r)=1-\frac{2M}{r}+\beta r-\frac{\Lambda r^2}{3}$, where $\beta=\alpha/d$ is a real constant.  For the range $R>> \Lambda$ and $R/R_0 >> 2/\alpha$ the action reduces to
\begin{equation}
f(R)= R+R_0 \ln{\left(\frac{R}{R_c}  \right)}~,
\end{equation}
 and satisfies the solar system range of $r << d$. On the other hand, at large scales, for $\alpha << 1$ and $R \sim R_0 \sim \Lambda$, the action tends to $f(R) \sim R+ \Lambda$, and agrees with cosmological observations. For metric (\ref{metric}) the curvature scalar is given by
\begin{equation}
R=\frac{6 \beta}{r}-4 \Lambda \,,
\end{equation}
 and the action tends to a constant value asymptotically
\begin{equation}
f(R) \approx \left( -3 \Lambda -\frac{3 R_0 \alpha \Lambda \ln{(- 3\Lambda/R_c)}}{2 R_0 - 4 \alpha \Lambda} \right) + \mathcal{O} (r^{-1})~.
\end{equation}
It is worth mentioning that the solution in both regimes was obtained as perturbation around the Einstein-Hilbert action and it was shown that within this approximation the solutions are consistent with the modified gravity equations. Also, the parameters of the theory were constrained by using the pioneer data, the flat rotation curve of galaxies and the cosmic microwave background radiation and Supernova Type Ia gold sample data giving $\beta\approx 10^{-26}m^{-1}$, $\beta d\approx 10^{-6}$ and $\beta d<<10^{-3}$, respectively \cite{Saffari:2007zt}. However,
the spacecraft's heat loss is a more plausible explanation for the
anomaly than the cosmological one \cite{Turyshev:2012mc}.

\begin{figure}[h]
\begin{center}
\includegraphics[width=0.58\textwidth]{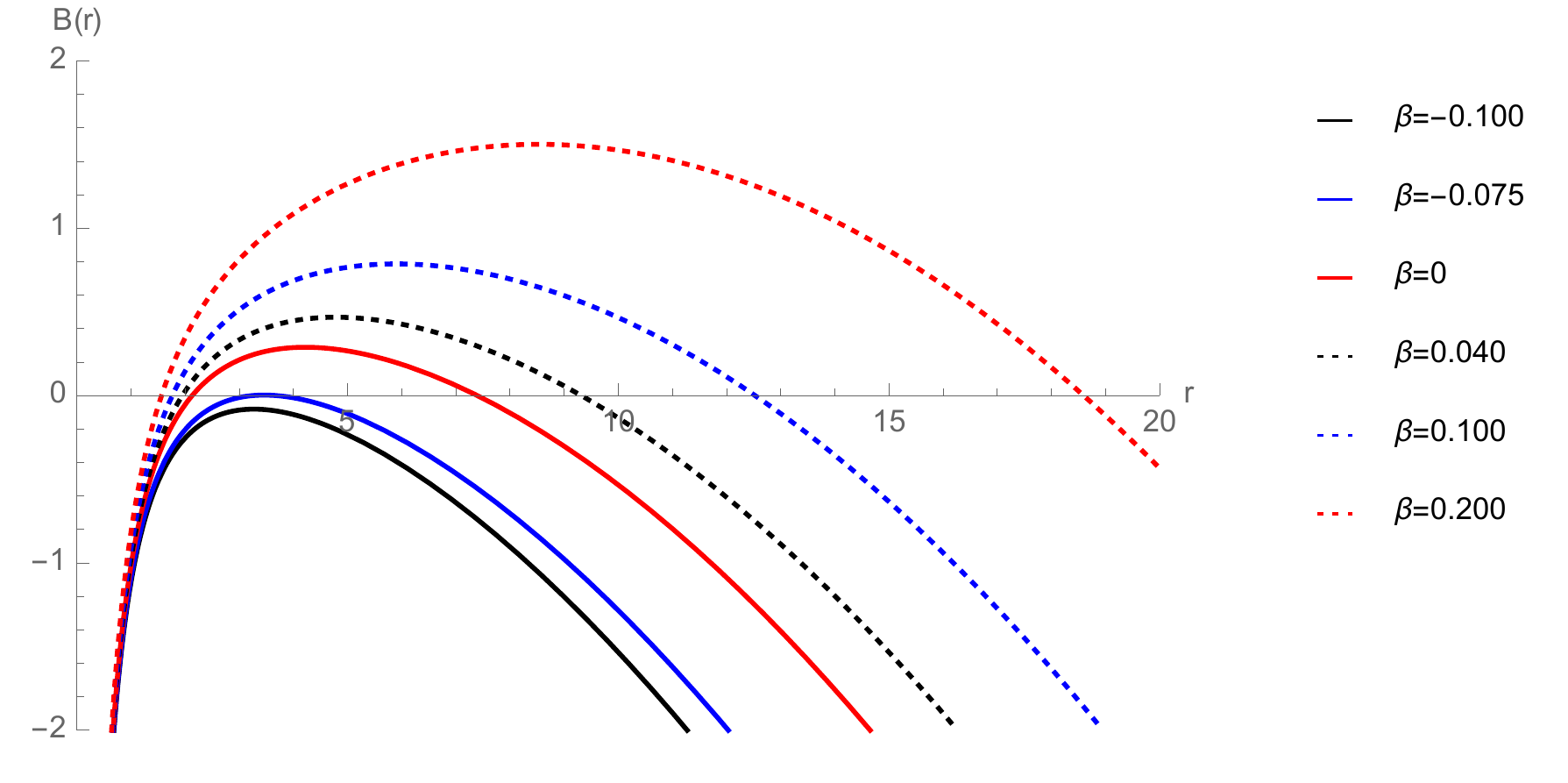}
\end{center}
\caption{The behaviour of $B(r)$ as a function of $r$ for different values of the parameters $\beta$ with $M=1$ and  $\Lambda=0.04$. }
\label{Function2}
\end{figure}

 Our analysis consider a range of values of $\beta$, such that the metric represents a black hole solution with an event horizon and a cosmological horizon as it shown in Fig. \ref{Function2}. In this figure for a fixed value of the cosmological constant,  we show the transition among black holes, near extremal ($r_H \approx r_{\Lambda}$) and naked singularity, when the $\beta$ parameter goes from positive to negative values. Notice that, for $\beta=0$, the spacetime is described by Schwarzschild-dS black hole. So, for $\beta$ positive (negative) the cosmological horizon radius is larger (less) than Schwarzschild-de Sitter black hole, while  for $\beta$ positive (negative) the event horizon radius is less (larger) than Schwarzschild-dS black hole, when the spacetime describes black hole solutions.

\section{Scalar perturbations}
\label{QNM}

In order to obtain the QNMs of scalar perturbations in the background of the metric (\ref{metric})
we consider the Klein-Gordon equation
\begin{equation}
\frac{1}{\sqrt{-g}}\partial _{\mu }\left( \sqrt{-g}g^{\mu \nu }\partial_{\nu } \varphi \right) =-m^{2}\varphi \,,  \label{KGNM}
\end{equation}%
with suitable boundary conditions for a black hole geometry. In the above expression $m$ is the mass
of the scalar field $\varphi $. Now, by means of the following ansatz
\begin{equation}
\varphi =e^{-i\omega t} R(r) Y(\Omega) \,,\label{wave}
\end{equation}%
the Klein-Gordon equation reduces to
\begin{equation}
\frac{1}{r^2}\frac{d}{dr}\left(r^2 B(r)\frac{dR}{dr}\right)+\left(\frac{\omega^2}{B(r)}-\frac{\ell (\ell+1)}{r^2}-m^{2}\right) R(r)=0\,, \label{radial}
\end{equation}%
where $\ell=0,1,2,...$ corresponds to the eigenvalue of the Laplacian on the two-sphere.
Now, defining $R(r)=\frac{F(r)}{r}$
and by using the tortoise coordinate $r^*$ given by
$dr^*=\frac{dr}{B(r)}$,
 the Klein-Gordon equation can be written as
  \begin{equation}\label{ggg}
 \frac{d^{2}F(r^*)}{dr^{*2}}-V_{eff}(r)F(r^*)=-\omega^{2}F(r^*)\,,
 \end{equation}
that corresponds to a one-dimensional Schr\"{o}dinger-like equation with an effective potential $V_{eff}(r)$, parametrized as $V_{eff}(r^*)$ and given by
  \begin{equation}\label{pot}
 V_{eff}(r)=\frac{B(r)}{r^2} \left(\ell (\ell+1) +  m^2 r^2+B^\prime(r)r\right)\,,
 \end{equation}
and substituting the metric  function $B(r)$ in the  effective potential in Eq. (\ref{pot}) we obtain
\begin{eqnarray}
\nonumber V_{eff}(r)&=&\frac{1}{3} \left(3 \beta ^2-\ell (\ell+1) \Lambda -2\Lambda +3m^2\right)+\frac{3\beta  \ell (\ell+1)+3\beta -6 m^2 M+2 \Lambda  M}{3 r}+\\
&&\frac{1}{9} r^2 \left(2 \Lambda ^2-3 \Lambda  m^2\right)+r \left(\beta  m^2-\beta  \Lambda \right)-\frac{4 M^2}{r^4}+\frac{-2\ell(\ell+1) M+2 M}{r^3}+\frac{\ell (\ell+1)}{r^2}\,.
\end{eqnarray}
Now, by considering $m=m_c$, as the value of the mass that cancels the divergence of order $r^2$, i.e, $m_c=\sqrt{2\Lambda/3}$,  and $\ell=0$, the effective potential is
\begin{equation}
    V_{eff}(r)=\beta ^2-\frac{4 M^2}{r^4}+\frac{2 M}{r^3}-\frac{2 \Lambda  M-3 \beta }{3 r}-\frac{\beta  \Lambda  r}{3}\,.
\end{equation}
So, we can note that for $\beta$ negative (positive) the potential diverges positive (negative) at infinity, and in principle the existence of a critical mass
could not be present in this case, due to the arguments given in \cite{Aragon:2020tvq}.\\

Now, in order to solve numerically the differential equation (\ref{radial}) by using the pseudospectral Chebyshev method \cite{Boyd},
it is convenient to perform the change of variable $y=(r-r_H)/(r_{\Lambda}-r_H)$. Thus,
the values of the radial coordinate are limited to the range $[0,1]$, and
the radial equation (\ref{radial}) can be written as
\begin{equation} \label{rad}
B(y) R''(y) + \left( \frac{2 \left(    r_{\Lambda}- r_H  \right) B(y)}{r_H+\left( r_{\Lambda}-r_H \right) y } + B'(y) \right) R'(y)+ \left( r_{\Lambda}-r_H  \right)^2 \left( \frac{\omega^2}{B(y)}- \frac{ \ell(\ell+1)}{\left( r_H + \left( r_{\Lambda}-r_H \right)y \right)^2} -m^2  \right) R(y)=0\,.
\end{equation}
Also, in order to propose an ansatz for $R(y)$ we consider its behavior in the vicinity of the horizon (y $\rightarrow$ 0)
\begin{equation}
R(y)=C_1 e^{-\frac{i \omega \left( r_{\Lambda}-r_H \right)}{B'(0)} \ln{y}}+C_2 e^{\frac{i \omega \left( r_{\Lambda}-r_H \right)}{B'(0)} \ln{y}} \,,
\end{equation}
and at the cosmological horizon ($y=1$)
\begin{equation}
R(y)= D_1 e^{-\frac{i \omega \left( r_{\Lambda}-r_H \right)}{B'(1)} \ln{(1-y)}}+D_2 e^{\frac{i \omega \left( r_{\Lambda}-r_H \right)}{B'(1)} \ln{(1-y)}}  \,.
\end{equation}
So, an ansatz that satisfy the requirements of only ingoing waves at the event horizon and at the cosmological horizon is
\begin{equation}
R(y)= e^{-\frac{i \omega \left( r_{\Lambda}-r_H \right)}{B'(0)} \ln{y}} e^{\frac{i \omega  \left( r_{\Lambda}-r_H \right)  }{B'(1)} \ln{(1-y)}} F(y) \,.
\end{equation}
Then, by inserting the above ansatz for $R(y)$ in Eq. (\ref{rad}), it is possible to obtain an equation for the function $F(y)$. The solution for the function $F(y)$ is assumed to be a finite linear combination of the Chebyshev polynomials, then it is inserted in the differential equation for $F(y)$. Also, the interval $[0,1]$ is discretized at the Chebyshev collocation points. Then, the differential equation is evaluated at each collocation point. So, a system of algebraic equations is obtained, and it corresponds to a generalized eigenvalue problem, which is solved numerically to obtain the QNFs. It is worth mentioning that, for $\beta=0$, the spacetime is described by the Schwarzschild-de Sitter black hole. The complex 
 family of QNMs for this geometry was determined in  \cite{Zhidenko:2003wq} by using the WKB and P\"oschl-Teller method. Also, it was shown that the frequencies all have a negative imaginary part, which means that the propagation of scalar field is stable in this background, and the presence of the cosmological constant leads to decrease of the real oscillation frequency and to a slower decay. High overtones were studied in  \cite{Konoplya:2004uk}. Also,  a family of purely imaginary  modes was reported in \cite{Jansen:2017oag} and an analysis of the photon sphere (PS) modes and de Sitter (dS) modes was recently performed in Ref. \cite{Aragon:2020tvq}.

\newpage

\subsection{The photon sphere family}

\subsubsection{Photon sphere modes}

The photon sphere  QNMs, are represented by damped oscillations whose decay rate ($\omega_I$) is directly connected to the instability timescale of null geodesics at the photon sphere. The photon sphere is a spherical trapping region of space where gravity is strong enough that photons are forced to travel in unstable circular orbits around a black hole. This region has a strong pull in the control of decay of perturbations and the QNMs with large frequencies. For instance, the decay timescale is related to the instability timescale of null geodesics near the photon sphere. For asymptotically dS black holes we find a family that can be traced back to the photon sphere and refer to them as PS modes. The dominant modes of this family are approached in the eikonal limit, where $\ell\rightarrow\infty$, for the mass of the scalar field lower than a critical value, and can be very well approximated with the WKB method.  For this family of modes, higher angular momentum represents smaller decay rates, such that in the limit of interest, the most representative modes are those for which $\ell \rightarrow \infty$ \cite{Cardoso:2017soq,Destounis:2018qnb}.

\begin{figure}[h]
\begin{center}
\includegraphics[width=0.46\textwidth]{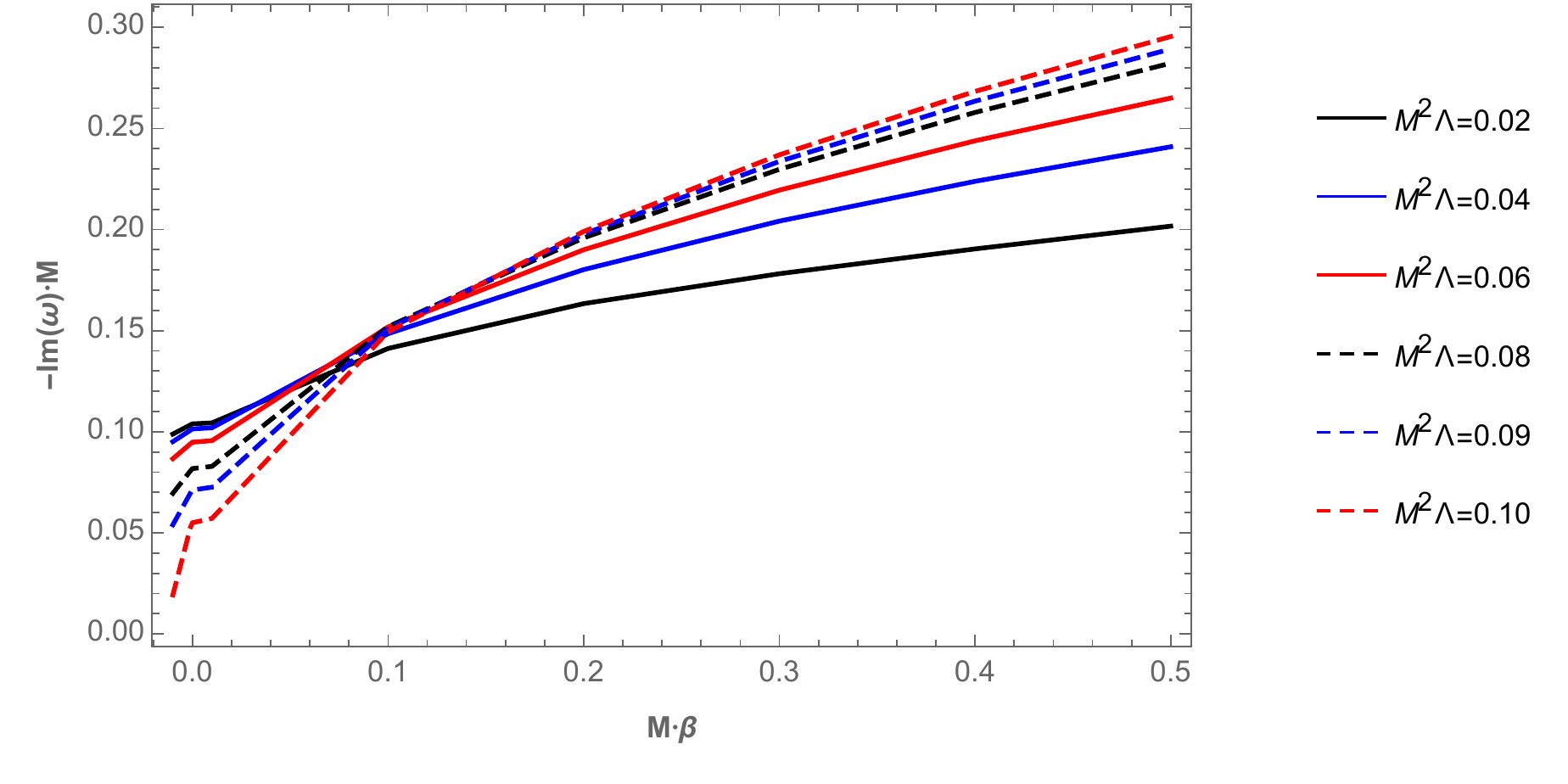}
\includegraphics[width=0.46\textwidth]{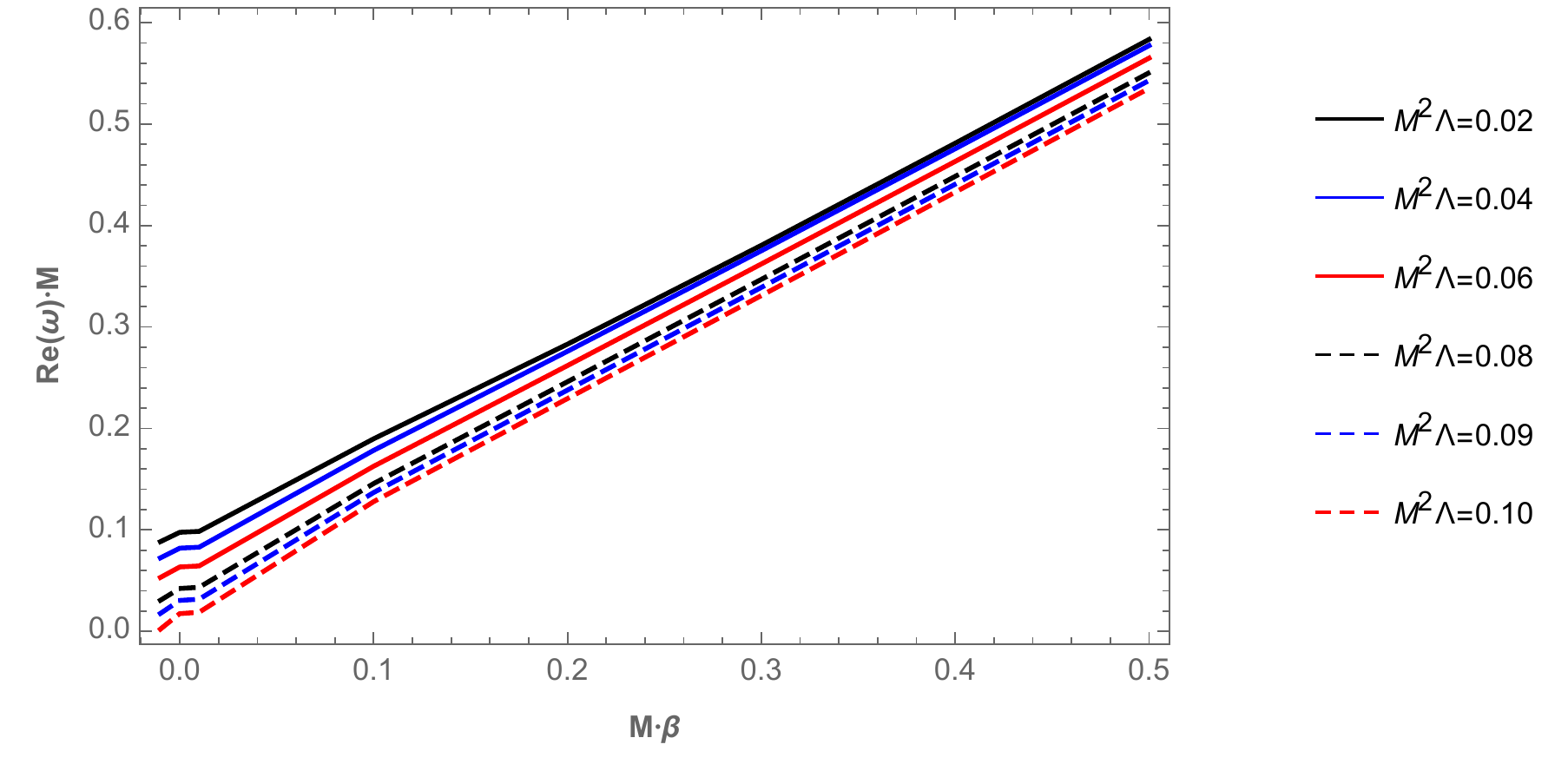}
\end{center}
\caption{The behavior of $Im(\omega) M$ (left panel) and $Re(\omega) M$ (right panel) for the fundamental QNF as a function of $M \beta$ for massless scalar field with $\ell=0$ and different values of the cosmological constant.}
\label{F5}
\end{figure}

\begin{figure}[h!]
\begin{center}
\includegraphics[width=0.46\textwidth]{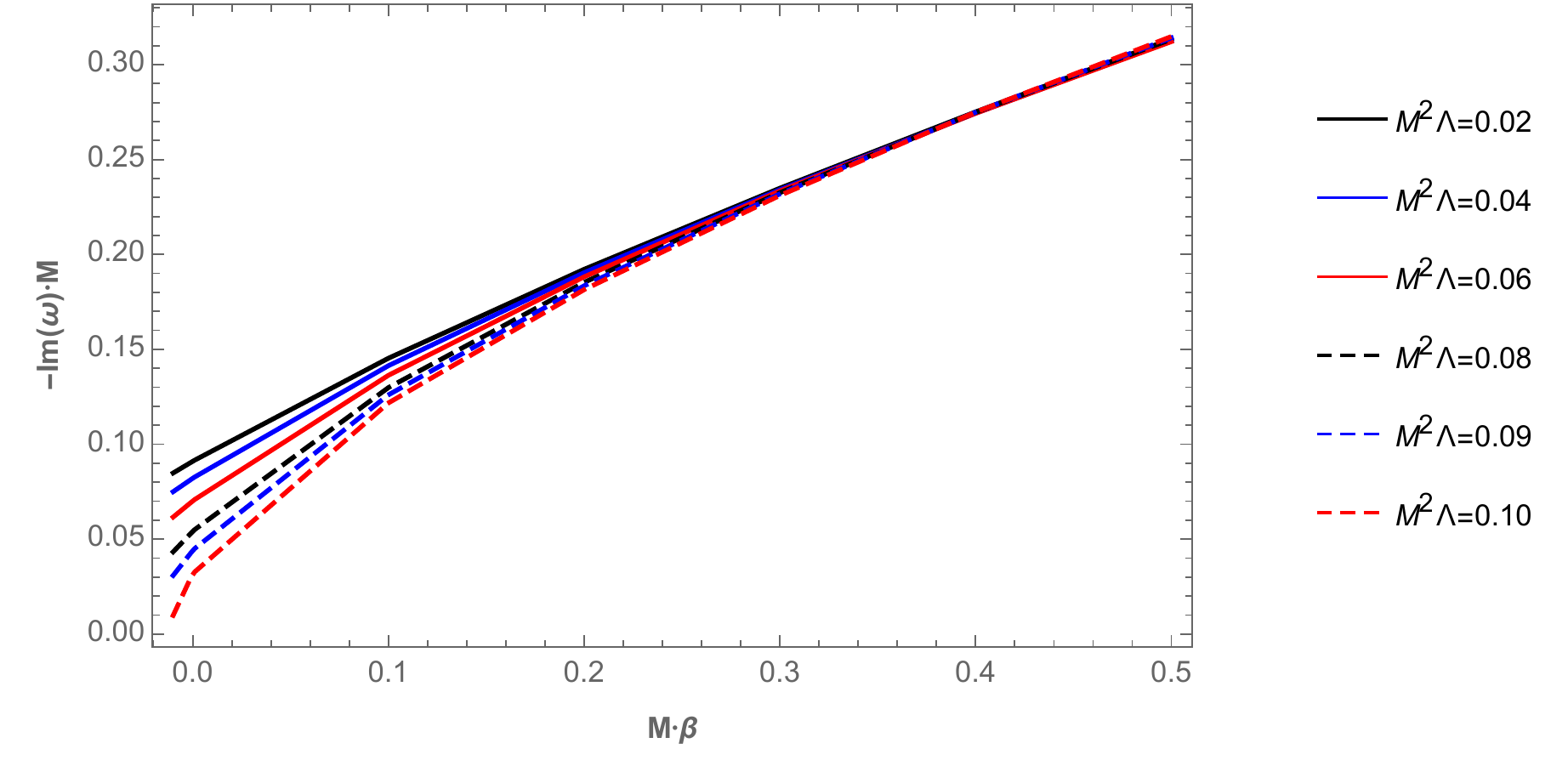}
\includegraphics[width=0.46\textwidth]{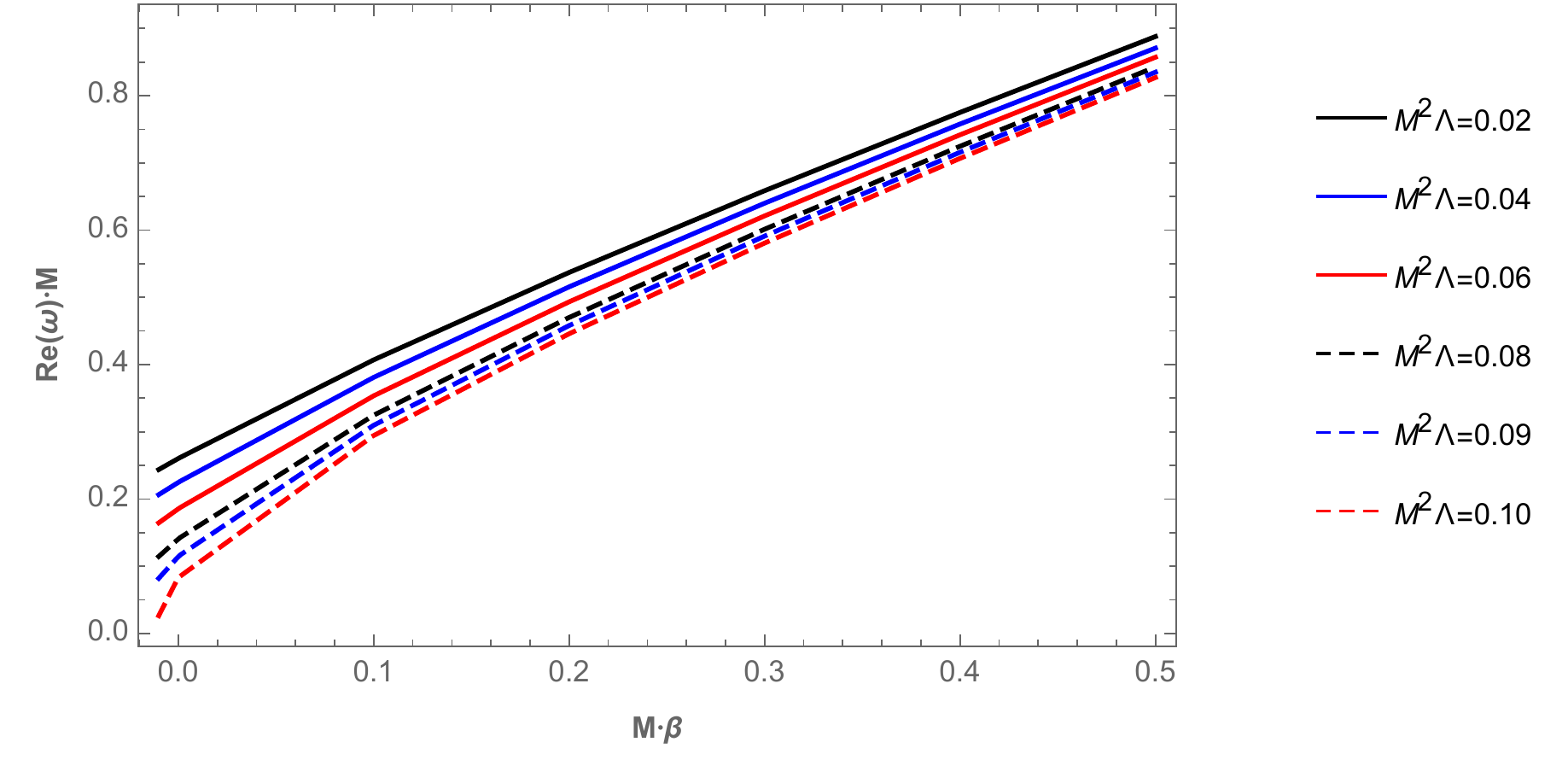}
\end{center}
\caption{The behaviour of $Im(\omega) M$ (left panel) and $Re(\omega) M$ (right panel) for the fundamental QNF as a function of $M \beta$ for massless scalar field with $\ell=1$ and different values of the cosmological constant.}
\label{F6}
\end{figure}

Now, in order to observe the behavior of photon sphere modes, we plot the behavior of the fundamental QNFs $n_{PS}=0$ \footnote{$n_{PS}$ corresponds to the overtone number of photon sphere modes} of massless scalar fields as a function of $M \beta$ with $\ell=0$, see Fig. \ref{F5}.
We can observe that for small values of $M \beta$, the decay rate decreases when the cosmological constant increases. However, for large $M \beta$, the behavior is opposite, i.e, the decay rate increases.
On the other hand, the frequency of the oscillation increases when $M \beta$ increases and decreases when the cosmological constant increases.
In Fig. \ref{F6} we consider $\ell=1$, note that the behavior of the QNFs is equivalent to the observed for $\ell=0$ and small $M \beta$; however,  there is not and inverted behavior, for the range of $M \beta$ values considered. Thus, when the cosmological constant increases the decay rate decreases, for small $M \beta$ and when $M \beta$ increases the decay rate converges to the same value and it does not depend on the value of the cosmological constant. On the other hand, the frequency of the oscillation increases when $M \beta$ increases and decreases when the cosmological constant increases, the same behavior was observed for $\ell=0$.

\subsubsection{Anomalous decay rate}

\begin{figure}[h]
\begin{center}
\includegraphics[width=0.46\textwidth]{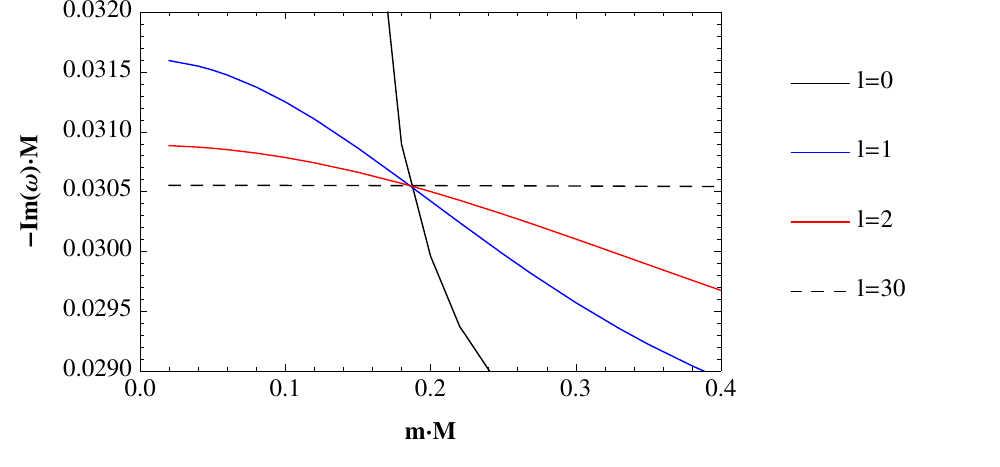}
\includegraphics[width=0.46\textwidth]{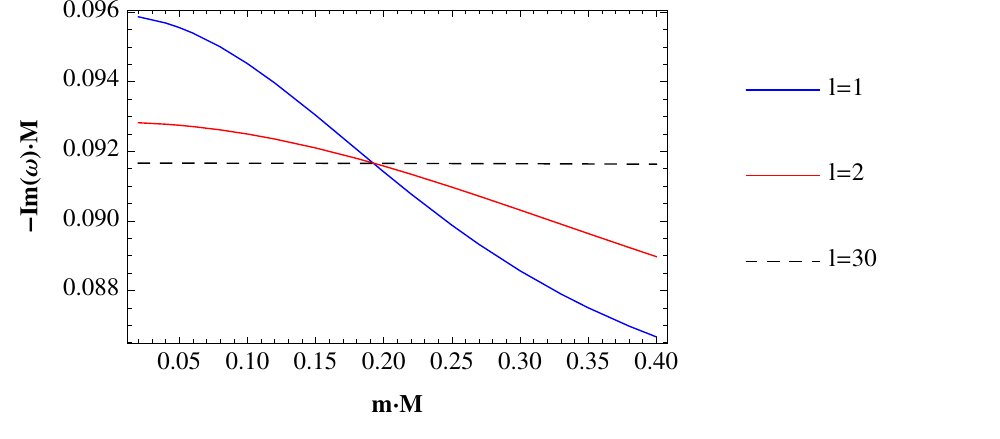}
\includegraphics[width=0.46\textwidth]{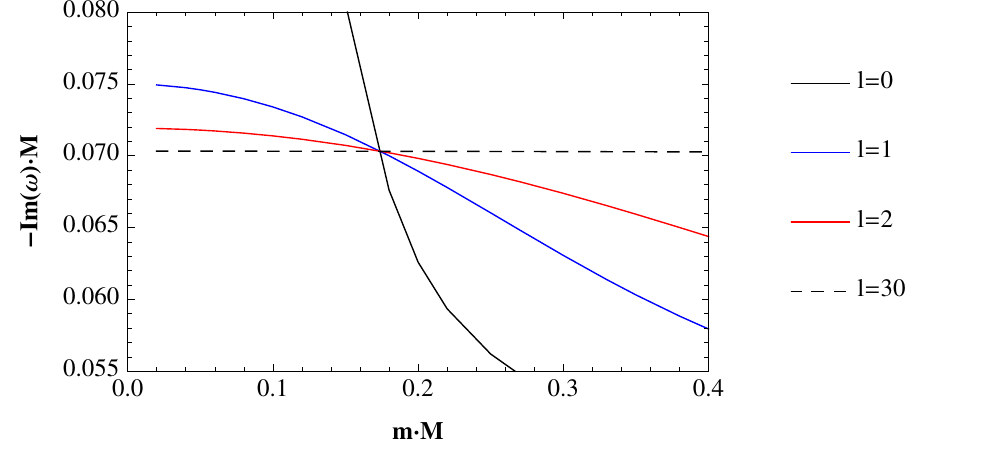}
\includegraphics[width=0.46\textwidth]{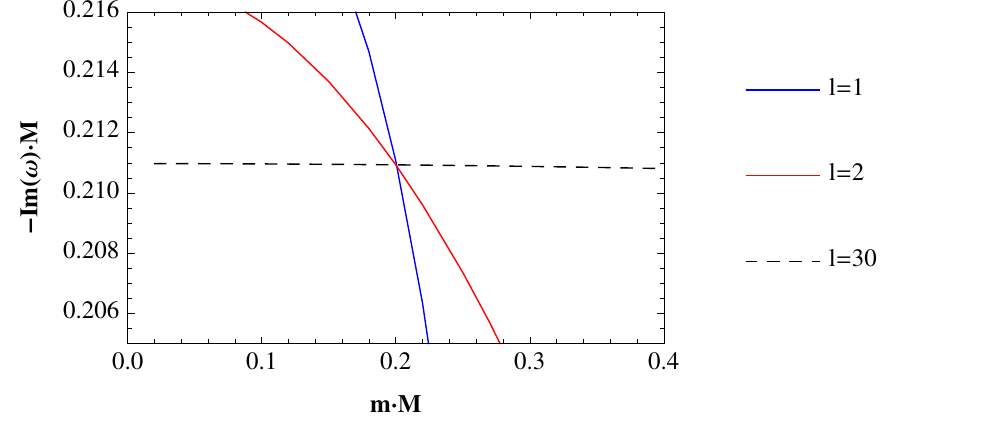}
\includegraphics[width=0.46\textwidth]{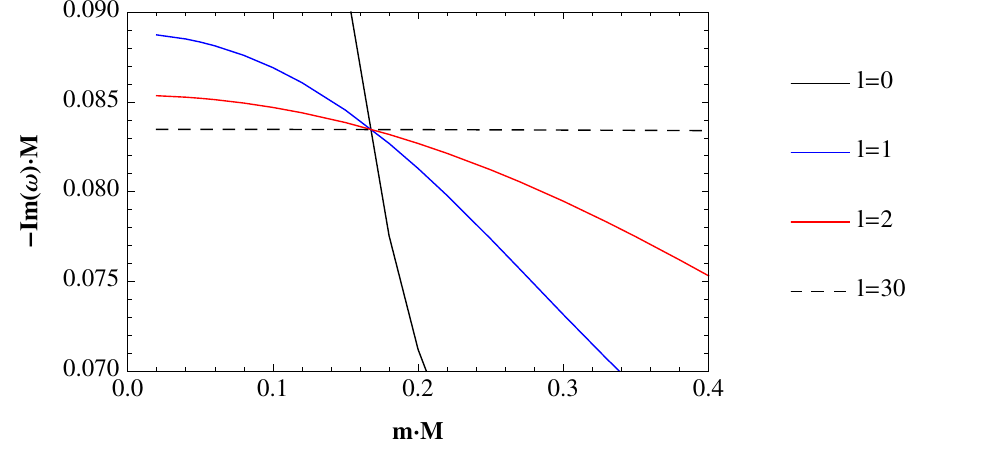}
\includegraphics[width=0.46\textwidth]{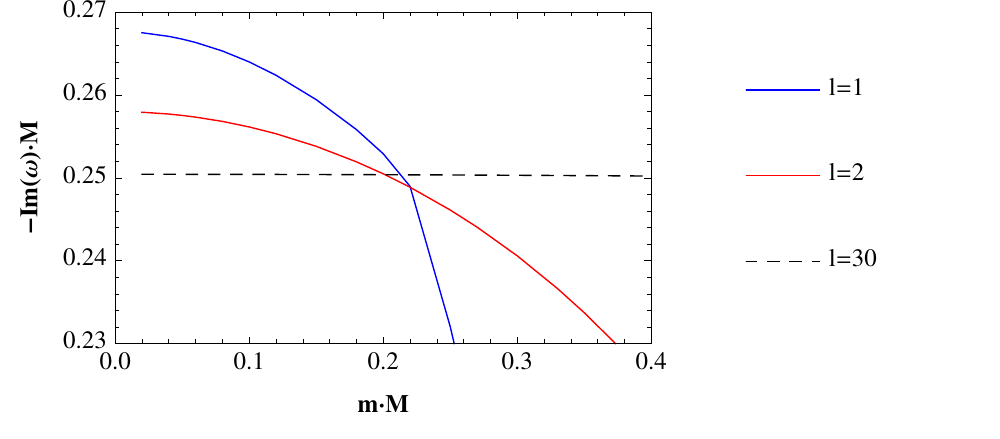}
\end{center}
\caption{The behavior of $-Im(\omega) M$
as a function of the scalar field mass for different values of the parameter $\ell=0,1,2,30$ and $n_{PS}=0$ (left panels), and $\ell=1,2,30$ and $n_{PS}=1$ (right panels) with $M^2 \Lambda=0.04$, and from top to bottom $M \beta= -0.06,-0.01$, and $0.01$, respectively.}
\label{F10}
\end{figure}

\begin{figure}[h]
\begin{center}
\includegraphics[width=0.46\textwidth]{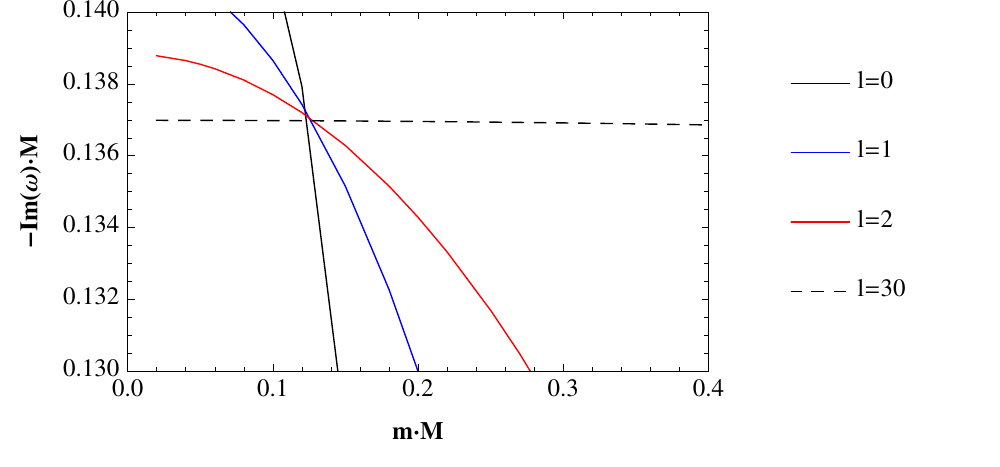}
\includegraphics[width=0.46\textwidth]{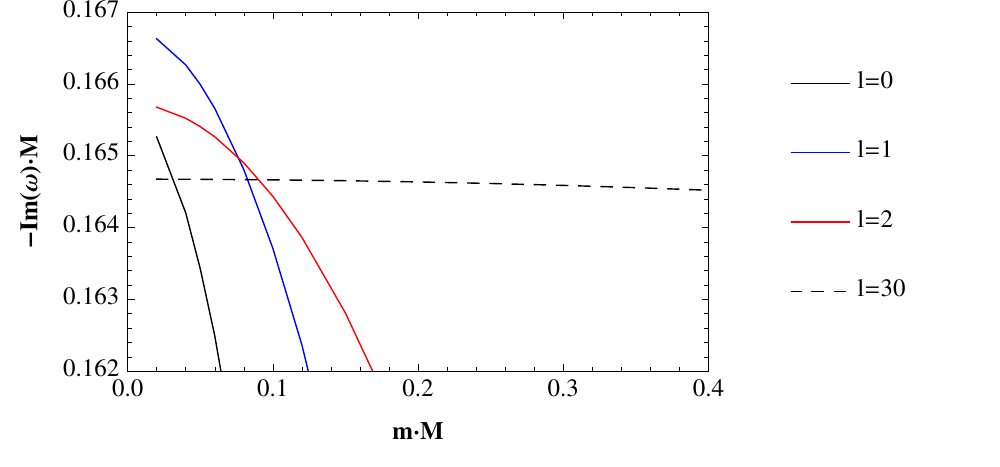}
\includegraphics[width=0.46\textwidth]{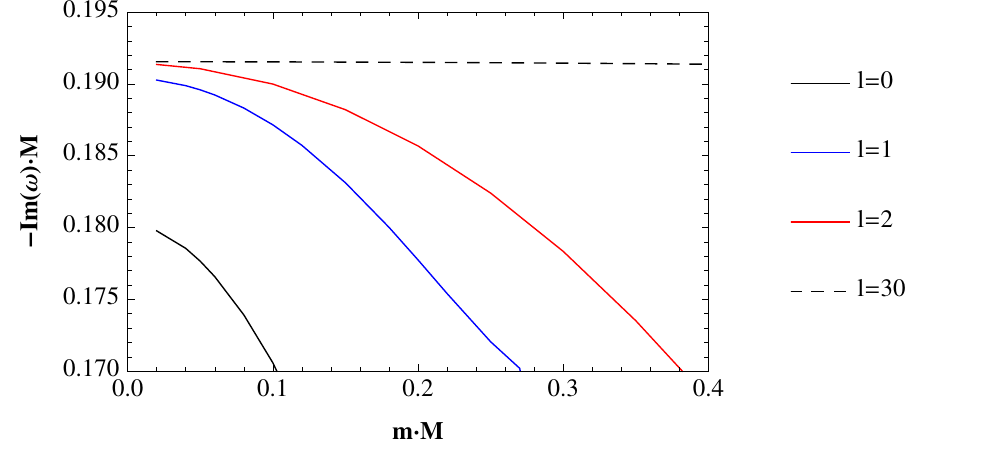}
\end{center}
\caption{The behavior of $-Im(\omega) M$  as a function of the scalar field mass $m M$ for different values of the parameter $\ell=0,1,2,30$ and $n_{PS}=0$ with $M^2 \Lambda=0.04$, and $M \beta=0.1$ (top panel), $M \beta=0.15$ (center panel) and $M \beta=0.20$ (bottom panel).}
\label{F11}
\end{figure}

Now, we will study if in $f(R)$ modified gravity, in the presence of a cosmological constant, the photon sphere modes
present an anomalous behavior, i.e the longest-lived modes are the ones with higher angular number, and also the presence of a critical mass, where beyond this value the behavior mentioned is inverted.
So, we plot the behavior of $-Im(\omega) M$
as a function of the scalar field mass for different values of the parameters $M\beta$, $\ell$ and $n_{PS}=0$ (left panels), and $n_{PS}=1$  with $\ell>n_{PS}$ (right panels)
\footnote{We have left outside the case for $\ell=0$, because the imaginary part of the QNFs exhibits a behavior other than $\ell>0$.
In general, this different behavior occurs for $\ell < n_{PS}$.},
for a fixed values of the black hole mass and the cosmological constant, see Fig. \ref{F10}. Also, we consider larger values of the parameter $M \beta$ with $n_{PS}=0$, see Fig. \ref{F11}.
We can observe that
the anomalous behavior along with the critical mass can be present for a certain range of values of the parameter $M \beta$. See, Fig. \ref{F10},
where the anomalous behavior occurs, and the critical mass appears, for $M \beta=-0.06, -0.01$, and $0.01$; and see Fig. \ref{F11},  where the anomalous behavior does occur for $M \beta=0.10$  and it does not occur and the critical mass does not appear, for $M \beta=0.15$, and $0.20$.
Note that, for $M \beta=0.20$, the longest-lived modes always are the ones with smaller angular number, but for $M \beta=0.15$, we cannot say the same.

As mentioned,
when $M \beta$ increases the metric function could represents naked singularity, extremal black hole, near extremal black hole and black hole solutions.
So, one of the effects of $M \beta$ is tuning the different cases, which have an important effect in the anomalous behavior, that is, for the near extremal case, see Fig. \ref{F10} top panel, the value of the critical mass is the same for the overtone numbers $n_{PS}=0$ and $n_{PS}=1$, which does not occur for the other case where the value of the critical mass increases when the overtone number increases. Also, we can observe that when the $M \beta$ parameter increases the critical mass is shifted by the effect of the $M \beta$ parameter, decreasing.  So,  there is a critical value of $M \beta$ where the critical mass is zero thereby there is not an anomalous behavior in the QNMs for larger values of $M \beta$. 
To explain this effect, we could say that the deviation of Schwarzschild-dS black hole, where the anomalous behavior and the critical mass have been observed, is given by the parameter $M \beta$ which also appears in the metric function. So for small values of this parameter, i.e small deviations Schwarzschild-dS black hole, the anomalous behavior occurs and the critical mass appears.
However, for larger values of the parameter $M \beta$ i.e large deviations from Schwarzschild-dS black hole, the QNMs have a different behavior, where  either the anomalous behavior and the critical mass
are not present, this cut off point {\bf{could}} numerically occurs for $M \beta\approx 0.10$, see Fig. \ref{F5}, where the QNFs present an inverted behavior with respect to $M \beta$ for massless scalar field.

\clearpage

\subsubsection{Analysis using the WKB method}
\label{WKBJ}

\begin{figure}[h]
\begin{center}
\includegraphics[width=0.46\textwidth]{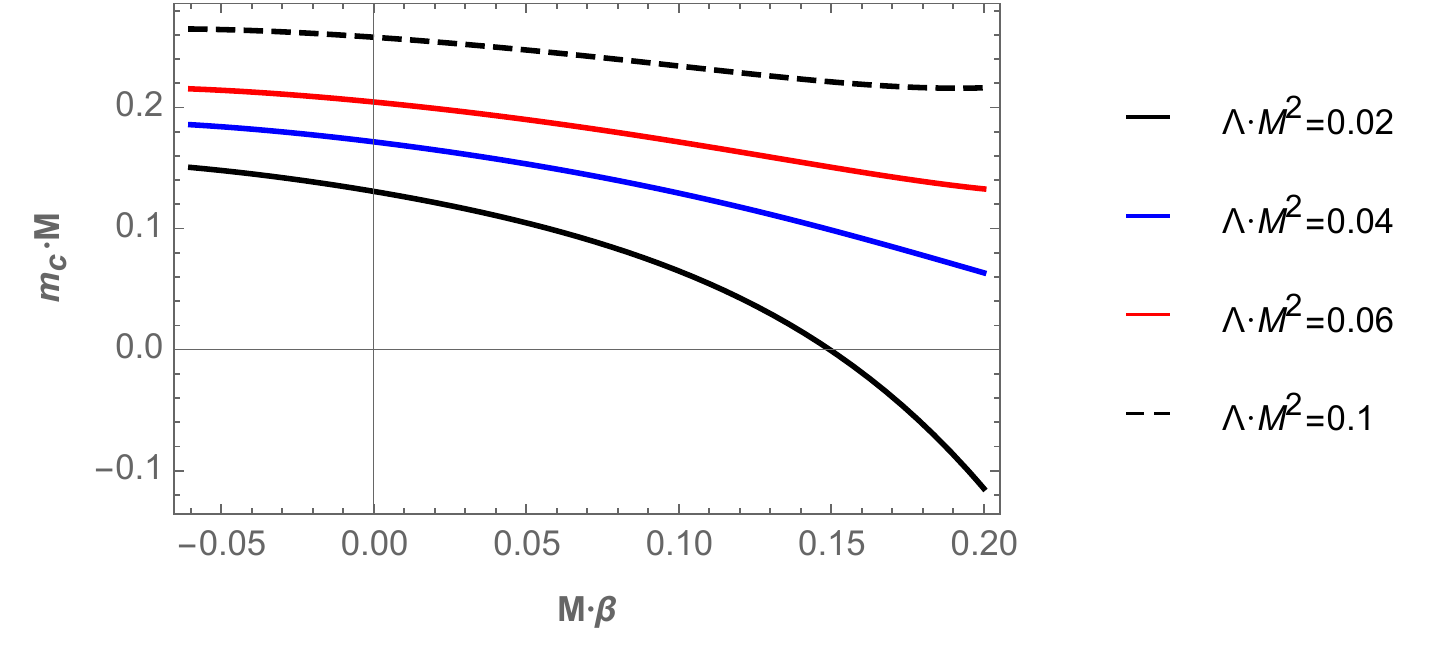}
\end{center}
\caption{The behavior of the $m_c M$ as a function of $M \beta$ for different values of $\Lambda M^2$ with $n_{PS}=0$.}
\label{F8}
\end{figure}

In this section, we use the method based on Wentzel-Kramers-Brillouin (WKB) in order to get some analytical insight of the behavior of the QNFs in the eikonal limit $\ell \rightarrow \infty$   \cite{Mashhoon,Schutz:1985zz,Iyer:1986np,Konoplya:2003ii,Matyjasek:2017psv,Konoplya:2019hlu}.
It is worth mentioning that this method can be used for effective potentials which have the form of potential barriers that approach to a constant value at the horizon and spatial infinity \cite{Konoplya:2011qq}.
The method consider
the behavior of the effective potential near its maximum value $r^*_{max}$.  So, by using the Taylor series expansion, the potential around its maximum is given by
\begin{equation}
V(r^*)= V(r^*_{max})+ \sum_{i=2}^{i=\infty} \frac{V^{(i)}}{i!} (r^*-r^*_{max})^{i} \,.
\end{equation}
where
\begin{equation}
V^{(i)}= \frac{d^{i}}{d r^{*i}}V(r^*)|_{r^*=r^*_{max}}\,,
\end{equation}
corresponds to the $i$-th derivative of the potential with respect to $r^*$ evaluated at the maximum of the potential $r^*_{max}$. Using the WKB approximation up to 6th order the QNFs are given by the following expression \cite{Hatsuda:2019eoj}
\begin{eqnarray} \label{ff}
\omega^2 &=& V(r^*_{max})  -2 i U \,,
\end{eqnarray}
where
\begin{eqnarray}
\notag U &=&  N\sqrt{-V^{(2)}/2}+\frac{i}{64} \left( -\frac{1}{9}\frac{V^{(3)2}}{V^{(2)2}} (7+60N^2)+\frac{V^{(4)}}{V^{(2)}}(1+4 N^2) \right)  \\
\notag && +\frac{N}{2^{3/2} 288} \Bigg( \frac{5}{24} \frac{V^{(3)4}}{(-V^{(2)})^{9/2}} (77+188N^2) +\frac{3}{4} \frac{V^{(3)2} V^{(4)}}{(-V^{(2)})^{7/2}}(51+100N^2)\\
\notag &&  +\frac{1}{8} \frac{V^{(4)2}}{(-V^{(2)})^{5/2}}(67+68 N^2)+\frac{V^{(3)}V^{(5)}}{(-V^{(2)})^{5/2}}(19+28N^2)+\frac{V^{(6)}}{(-V^{(2)})^{3/2}} (5+4N^2)  \Bigg)\,,
\end{eqnarray}
and $N=n_{PS}+1/2$, with $n_{PS}=0,1,2,\dots$, is the overtone number.
Now, by  defining $L^2= \ell (\ell+1)$, we find that for large values of $\ell$, the maximum of the potential is approximately at
\begin{equation}
\notag r_{max} \approx  r_0 + \frac{1}{L^2} r_1 + \mathcal{O}(L^{-4}) \,,
\end{equation}
where
\begin{eqnarray}
\notag r_0 & \approx & 3M - \frac{9 M^2}{2} \beta + \frac{27 M^3}{2} \beta^2 -\frac{405 M^4}{8}\beta^3 + \mathcal{O} (\beta^4)\,,    \\
\notag r_1 & \approx & -\frac{1}{3}M (-1+9M^2(3m^2-2 \Lambda)) (-1+9 \Lambda M^2) + \frac{1}{2} M^2 (-1+9 M^2 (3 m^2-2 \Lambda)) (-1+63 M^2 \Lambda ) \beta + \\
\notag && \frac{9}{4} M^3 (2+9 M^2 \Lambda (13+216 M^2 \Lambda) -27 m^2 (M^2+108 M^4 \Lambda)) \beta^2 +\frac{9}{8}M^4(-35+9M^2(-2 \Lambda (98+1485 \Lambda M^2)+  \\
\notag && m^2 (84+4455 \Lambda M^2))) \beta^3 + \mathcal{O} (\beta^4)\,,
\end{eqnarray}
and
\begin{equation} \label{coa}
\notag V(r^*_{max}) \approx V_0 L^2 +V_1 + \mathcal{O}(L^{-2}) \,,
\end{equation}
where
\begin{eqnarray}
 \notag V_0 & \approx &    \left( \frac{1}{27 M^2}-\frac{\Lambda}{3} \right) +\frac{\beta}{3M}+\frac{\beta^2}{4} -\frac{M \beta^3}{4} + \mathcal{O}(\beta^4)\,, \\
 \notag V_1 & \approx &  - \frac{(2+9 M^2(3m^2-2\Lambda ))(-1+9 \Lambda M^2)}{81 M^2} + \left(  \frac{10}{27 M} +M \left( 2m^2 -\frac{8 \Lambda}{3} \right) +3 M^3 (3m^2-2 \Lambda) \Lambda \right) \beta +  \\
\notag && \left( \frac{23}{18} + \frac{1}{4} M^2 (-3m^2 (4+45 \Lambda M^2) + 2\Lambda (8+45 \Lambda M^2)) \right) \beta^2  + \frac{1}{4} M (-2 +3M^2 (3m^2(4+63 \Lambda M^2)-2 \Lambda (8+  \\
&&  63 \Lambda M^2))) \beta^3 + \mathcal{O} (\beta^4) \,.
\end{eqnarray}
The second derivative of the potential evaluated at $r^*_{max}$ yields the expression
\begin{equation}
V^{(2)}(r^*_{max}) \approx V^{(2)}_0 L^2 +V^{(2)}_1 + \mathcal{O}(L^{-2}) \,,
\end{equation}
where
\begin{eqnarray}
\notag V^{(2)}_0  & \approx & -\frac{2 (1- 9 \Lambda M^2)^2}{729 M^4} - \frac{2 ((-7 + 9 \Lambda M^2)(-1+9 \Lambda M^2))}{243 M^3} \beta + \left( -\frac{32}{81 M^2}+ \frac{13 \Lambda}{9} + \Lambda^2 M^2 \right)  \beta^2  \\
\notag && -\frac{(8+ 3 \Lambda M^2(2+9 \Lambda M^2))}{9M} \beta^3 + \mathcal{O} (\beta^4) \,, \\
\notag V^{(2)}_1 & \approx &  \frac{2(1-9 \Lambda M^2)^2 (-8+9M^2 (m^2 (6-135 \Lambda M^2)+2 \Lambda (-1+45 \Lambda M^2)))}{6561 M^4} + \\
\notag && \frac{2(-1+9 \Lambda M^2)(64+27M^2 (4 \Lambda-54 \Lambda^2 M^2(3+11 \Lambda M^2)+ m^2(-16 +9 \Lambda M^2 (23+99 \Lambda M^2))))}{2187 M^3} \beta +  \\
\notag && \frac{(-116+9M^2 (-3m^2 (-25 + 9 \Lambda M^2 (13+18 \Lambda M^2 (8+63 \Lambda M^2)))+ \Lambda (5 +27 \Lambda M^2 (14 +3 \Lambda M^2 (41 +252 \Lambda M^2)))))}{243 M^2} \beta ^2 \\
\notag && + \left( -\frac{344}{243 M} +4M (m^2 - \Lambda) +5M^3(9m^2-10 \Lambda)+6M^5 (126 m^2-107 \Lambda) \Lambda^2 +2106 M^7 (3 m^2-2 \Lambda) \Lambda^3 \right) \beta^3 + \mathcal{O} (\beta^4)\,,
\end{eqnarray}
and the higher derivatives of the potential evaluated at $r^*_{max}$ yields the expressions
\begin{eqnarray}
\notag V^{(3)}(r^*_{max}) \approx && \Bigg( -\frac{4(-1+9 \Lambda M^2)^3}{6561 M^5} +\frac{4(1-9 \Lambda M^2)^2}{243M^4} \beta +\frac{(-13+9 \Lambda M^2)(-1+9 \Lambda M^2)}{81 M^3} \beta^2 +\frac{53-9 \Lambda M^2 (16+9 \Lambda M^2)}{81 M^2} \beta^3  \\ \notag &&  + \mathcal{O}(\beta^4)   \Bigg)L^2 + \mathcal{O}(L^0)\,, \\
\notag V^{(4)}(r^*_{max}) \approx \notag && \Bigg( -\frac{16 (-1+ 9 \Lambda M^2)^3}{19683 M^6} -\frac{16 (1-9 \Lambda M^2)^2 (-11+18 \Lambda M^2)}{6561 M^5} \beta + \frac{4 (-1+9 \Lambda M^2) (-7+27\Lambda M^2)}{81 M^4} \beta^2 + \\
\notag && \frac{4(131 +45 \Lambda M^2 (-20 +9 \Lambda M^2))}{243 M^3} \beta^3 +\mathcal{O}(\beta^4)   \Bigg) L^2 + \mathcal{O} (L^0)\,,  \\
\notag V^{(5)}(r^*_{max}) \approx \notag &&  \Bigg( -\frac{40(1-9 \Lambda M^2) ^4}{59049 M^7} -\frac{40 (13-9 \Lambda M^2)(1-9 \Lambda M^2)^3}{19683 M^6} \beta + \frac{20 (1- 9 \Lambda M^2)^2 (-137+9\Lambda M^2(28+9 \Lambda M^2 )) }{6561 M^5}  \beta^2 \\
\notag && + \frac{20(1 -9 \Lambda M^2)(-371+9\Lambda M^2 (149+9 \Lambda M^2 (5+9 \Lambda M^2)))}{2187 M^4} \beta^3 + \mathcal{O}(\beta^4) \Bigg)L^2 +\mathcal{O}(L^0)\,,  \\
\notag V^{(6)}(r^*_{max}) \approx && \Bigg(  -\frac{16(1-9\Lambda M^2)^4 (4+15 \Lambda M^2)}{177147 M^8} +  \\
\notag && \frac{64(-1+9 \Lambda M^2)^3(5+6 \Lambda M^2)}{19683 M^7} \beta -\frac{4(1-9\Lambda M^2)^2 (1517 +9\Lambda M^2 (-302+81 \Lambda M^2))}{19683 M^6} \beta^2  \\
\notag && +\frac{4(-1+9\Lambda M^2)(15655 +9 \Lambda M^2 (-9161 + 9 \Lambda M^2 (841+81 \Lambda M^2))}{19683 M^5} \beta^3 + \mathcal{O}(\beta^4) \Bigg) L^2 + \mathcal{O}(L^0)\,.
\end{eqnarray}
Then, using the expressions above with Eq. (\ref{ff}), we obtain the following analytical expression for the QNFs for large values of $\ell$ and small values of $M \beta$
\begin{equation}
\label{omegawkb}
  \omega \approx \omega_{-1} L + \omega_0 + \omega_1 L^{-1} + \omega_2 L^{-2} + \mathcal{O}(L^{-3}) \,,
\end{equation}
where
\begin{eqnarray}
\notag \omega_{-1} & \approx & \frac{\sqrt{1-9 \Lambda M^2}}{3 \sqrt{3}M} + \frac{\sqrt{3}}{2 \sqrt{1-9 \Lambda M^2}} \beta - \frac{3 \sqrt{3} M (2+9\Lambda M^2)}{8(1-9 \Lambda M^2)^{3/2}} \beta^2  \\
\notag && +\frac{3 \sqrt{3} M^2\sqrt{1-9 \Lambda M^2}(-16+9 \Lambda M^2 (-13+18 \Lambda M^2))}{16(-1+9 \Lambda M^2)^3} \beta^3 + \mathcal{O}(\beta^4) \,, \\
\notag \omega_0 & \approx & -i\frac{\sqrt{1-9 \Lambda M^2}}{6 \sqrt{3} M } - i \frac{4-9 \Lambda M^2}{4 \sqrt{3} \sqrt{1-9 \Lambda M^2 }}\beta + i\frac{3\sqrt{3} M (1+9 \Lambda M^2 (1+9 \Lambda M^2))}{16(1-9 \Lambda M^2)^{3/2}} \beta^2  \\
\notag &&+i\frac{3 \sqrt{3}M^2(-8+9\Lambda M^2 (-13+9 \Lambda M^2 (-13+63 \Lambda M^2)))}{32(1-9 \Lambda M^2)^{5/2}} \beta^3 + \mathcal{O}(\beta^4)\,,  \\
\notag \omega_1 & \approx & -\frac{\sqrt{1-9 \Lambda M^2} (-34+9 M^2 (-108 m^2+61 \Lambda))}{648 M \sqrt{3}} + \frac{314-45 \Lambda M^2(59+180 \Lambda M^2)+972m^2(M^2+18 \Lambda M^4)}{432 \sqrt{3} \sqrt{1-9 \Lambda M^2}} \beta   \\
\notag && - \frac{-710 M-5832 M^7 (135 m^2-46 \Lambda) \Lambda^2 +108 M^3 (72m^2+5 \Lambda) +243 \Lambda M^5 (396 m^2+ 13 \Lambda)}{576 \sqrt{3} (1-9 \Lambda M^2)^{3/2}} \beta^2  + \\
\notag && M^2 \Big(-1262 +9 M^2 (\Lambda (-466+27 \Lambda M^2 (335 + 6\Lambda M^2 (301-2736 \Lambda M^2)))+324 m^2 (10+3 \Lambda M^2 (31+   \\
\notag && 18 \Lambda M^2 (-31 +126 \Lambda M^2)))) \Big) \beta^3 \Big/ \Big(384 \sqrt{3} (1-9 \Lambda M^2)^{5/2} \Big) + \mathcal{O}(\beta^4)\,, \\
\notag  \omega_2 & \approx &  -i \frac{(137+45 M^2 (-648 m^2 +401 \Lambda))(1-9 \Lambda M^2)^{3/2}}{23328 M \sqrt{3}}\\
\notag && +i \frac{\sqrt{1-9 \Lambda M^2} (590+17496 m^2 M^2 (2+27 \Lambda M^2) - 9 \Lambda M^2 (3593+ 30807 \Lambda M^2))}{5184 \sqrt{3}} \beta \\
\notag && + i\frac{M (7949+9 M^2 (648 m^2 (-23+9 \Lambda M^2 (-11+711 \Lambda M^2))+ \Lambda (8714 -9 \Lambda M^2 (7858 +252747 \Lambda M^2))))}{6912 \sqrt{3} \sqrt{1- 9 \Lambda M^2}} \beta^2 \\
\notag && +i M^2 \Big( -23674+81M^2 (\Lambda (-2219+3 \Lambda M^2 (70514 +627708 \Lambda M^2 -8621073 \Lambda^2 M^4)) \\
\notag && +72 m^2 (154+27 \Lambda M^2 (-65+27 \Lambda M^2 (-119+977 \Lambda M^2)))) \Big) \beta^3 \Big/ \Big( 13824 \sqrt{3} (1-9 \Lambda M^2)^{3/2}  \Big) + \mathcal{O} (\beta^4)\,.
\end{eqnarray}
The term proportional to $1/L^2$ is zero at the value of the critical mass $m_c$, which is given by
\begin{eqnarray}\label{mc}
\notag m_c M  \approx && \sqrt{\frac{137+18045 M^2 \Lambda}{29160}}-\frac{(1886-4185M^2 \Lambda) M\beta}{60 \sqrt{10}\sqrt{137+18045 M^2 \Lambda}}-\frac{3(4634227+134256255 M^2 \Lambda+1091552625 M^4 \Lambda^2)M^2 \beta^2}{400 \sqrt{10} (137+18045 M^2 \Lambda)^{3/2}} \\
 &&  - \frac{9 (5019824246+76442598769 M^2 \Lambda-4248331450335 M^4 \Lambda^2 -85433253134175 M^6 \Lambda^3) M^3 \beta^3}{800 \sqrt{10} (137+18045 M^2 \Lambda)^{5/2}}+ \dots\,.
\end{eqnarray}
and it is valid for small values of $\beta$ and $n_{PS}=0$.  In all the above expressions we have performed a Taylor's expansion around $\beta=0$ in order to obtain expressions that will be easy to handle analytically.
For $\beta=0$ we recover the result of critical mass for Schwarzschild-dS black holes \cite{Aragon:2020tvq}, and for $\Lambda M^2 = 0.04$ and $n_{PS}=0$ we obtain from (\ref{mc}) the value $m_c M=0.186, 0.175, 0.168$, and $0.130$ for $\beta=-0.06, -0.01, 0.01$, and $0.10$, respectively, which agrees with the numerical results shown in Fig. \ref{F10}, thereby the value of the critical mass decreases when the beta parameter increases and it increases when $\Lambda M^2$ increases, see Fig. \ref{F8}. Also, note that for small values of $\Lambda M^2$ there is a range of values for $M \beta$ where $m_c M$ becomes negative. Also, the WKB method proposes a critical scalar field mass for $M \beta=0.20$, contrary to the observed via the pseudospectral Chebyshev method. However, the analysis performed with the WKB method is valid only for small values of $\beta$, because we have performed a Taylor's expansion around $\beta=0$ to obtain the formula (\ref{mc}).
Now, in order to check the correctness and accuracy of the pseudospectral Chebyshev method with respect to the analytical expression for the QNFs given by Eq. (\ref{omegawkb}), we show in Table \ref{QNMbeta}, the values obtained with both methods.
Also, we show the relative error, which is defined by
\begin{eqnarray}
\label{E}
\epsilon_{Re(\omega)} &=& \frac{\mid Re(\omega_1)\mid- \mid Re(\omega_0)\mid}{\mid Re(\omega_0) \mid}\cdot 100\%\,, \\
\epsilon_{Im(\omega)} &=&\frac{\mid Im(\omega_1) \mid- \mid Im(\omega_0)\mid}{\mid Im(\omega_0)\mid } \cdot 100\% \,,
\end{eqnarray}
where $\omega_1$ corresponds to the result from Eq. (\ref{omegawkb})
and $\omega_0$ denotes the result with the pseudospectral Chebyshev method.  We can observe that the error does not exceed $0.0008$ $\%$ in the imaginary part, and $0.0010$ $\%$ in the real part. Also, as it was observed, the frequencies all have a negative imaginary part, which means that the propagation of massive scalar fields is stable in this background.

\begin {table}[ht]
\caption {Fundamental quasinormal frequencies ($n_{PS} =
      0 $) for massive scalar fields with $\ell = 10, 20, $ and $30$ in the background of black hole solution of $f(R)$ gravity with $M^2 \Lambda = 0.04 $, and $M \beta=0.01$.}
\label {QNMbeta}\centering
\begin {tabular} { | c | c | c | c |c |}
\hline
\multicolumn{5}{ |c| }{$m M= 0.02$} \\ \hline
$\ell$ & $\text{Pseudospectral method} $ & $ \text{WKB}$ & $ \epsilon_{Re(\omega)} $ & $ \epsilon_{Im(\omega)} $ \\\hline
$10$ &
$1.72651001 - 0.08356503 i $ &
$1.72652389 - 0.08356574 i $ &
$0.0008 $ &
$0.0008$ \\
$20$ &
$3.37194090 - 0.08348702 i $ &
$3.37197238 - 0.08348760 i$ &
$0.0009 $ &
$0.0007$ \\
$30$ &
$5.01712051 - 0.08347182 i $ &
$5.01716774 - 0.08347239 i $ &
$0.0009$ &
$0.0007$ \\\hline
\multicolumn{5}{ |c| }{$m M= 0.15$} \\ \hline
$\ell$ & $\text{Pseudospectral method} $ & $ \text{WKB}$ & $ \epsilon_{Re(\omega)} $ & $ \epsilon_{Im(\omega)} $ \\\hline
$10$ &
$1.72802056 - 0.08348145 i $ &
$1.72803867 - 0.08348207 i $ &
$0.0010 $ &
$0.0007$ \\
$20$ &
$3.37271554 - 0.08346511 i $ &
$3.37274759 - 0.08346568 i $ &
$0.0010 $ &
$0.0007$ \\
$30$ &
$5.01764129 - 0.08346193 i $ &
$5.01768870 - 0.08346249 i $ &
$0.0009  $ &
$0.0007$ \\\hline
\multicolumn{5}{ |c| }{$m M= 0.30$} \\ \hline
$\ell$ & $\text{Pseudospectral method} $ & $ \text{WKB}$ & $ \epsilon_{Re(\omega)} $ & $ \epsilon_{Im(\omega)} $ \\\hline
$10$ &
$1.73263370 - 0.08322691 i $ &
$1.73266525 - 0.08322649 i $ &
$0.0018 $ &
$0.0005$ \\
$20$ &
$3.37508142 - 0.08339825 i  $ &
$3.37511532 - 0.08339874 i$ &
$0.0010$ &
$0.0006$ \\
$30$ &
$5.01923188 - 0.08343171 i $ &
$5.01927986 - 0.08343226 9 $ &
$0.0010$ &
$0.0007$ \\\hline
\end {tabular}
\end {table}

As we mentioned, the Taylor's series have been performed around $\beta=0$, so the WKB analysis is true for small values of $\beta$. In table \ref{QNMbetaE}, we show the error in the quasinormal frequency between the WKB and pseudospectral method, for different values of $mM$, and $M\beta$. Note that for $M\beta=0.20$ the error  is $18$ $\%$ in the imaginary part, and $31$ $\%$ in the real part, approximately, that mean a loss of robustness and diminished validity in the results. Also, we can observe that the error increases slightly, when $mM$ increases.

\begin {table}[ht]
\caption {Fundamental quasinormal frequencies ($n_{PS} =
      0 $) for massive scalar fields with $\ell = 30$ in the background of black hole solution of $f(R)$ gravity with $M^2 \Lambda = 0.04 $, and different values of $M \beta$.}
\label {QNMbetaE}\centering
\begin {tabular} { | c | c | c | c |c |}
\hline
\multicolumn{5}{ |c| }{$m M= 0.02$} \\ \hline
$M\beta$ & $\text{Pseudospectral\, method} $ & $ \text{WKB}$ & $ \epsilon_{Re(\omega)} $ & $ \epsilon_{Im(\omega)} $ \\\hline
$-0.06$ &
$2.08314131 - 0.03055262 i$ &
$2.25222758-0.03267338 i$ &
$8.11689$ &
$6.94134$ \\
$-0.01$ &
$4.35540688 - 0.07032092 i$ &
$4.35546389 - 0.07032160 i$ &
$0.00130895$ &
$0.000966995$ \\
$0.01$ &
$5.01712051 - 0.08347182 i$ &
$5.01716774 - 0.08347239 i$ &
$0.000941377$ &
$0.000682865$ \\
$0.10$ &
$7.42982610 - 0.13698770 i$ &
$7.70221565 - 0.14017416 i$ &
$3.66616$ &
$2.32609$ \\
$0.15$ &
$8.55684679 - 0.16467397 i$ &
$9.67805702 - 0.17771512 i$ &
$13.1031$ &
$7.91938$ \\
$0.20$ &
$9.59615293 - 0.19154767 i$ &
$12.58587755 - 0.22617900 i$ &
$31.1554$ &
$18.0797$ \\\hline
\multicolumn{5}{ |c| }{$m M= 0.15$} \\ \hline
$M\beta$ & $\text{Pseudospectral\, method} $ & $ \text{WKB}$ & $ \epsilon_{Re(\omega)} $ & $ \epsilon_{Im(\omega)} $ \\\hline
$-0.06$ &
$2.08341635 - 0.03055125 i$ &
$2.25253125 - 0.03267230 i$ &
$8.11719$ &
$6.9426$ \\
$-0.01$ &
$4.35588702 - 0.07031330 i$ &
$4.35594414 - 0.07031399 i$ &
$0.00131133$ &
$0.000981322$ \\
$0.01$ &
$5.01764129 - 0.08346193 i$ &
$5.01768870 - 0.08346249 i$ &
$0.000944866$ &
$0.000670965$ \\
$0.10$ &
$7.43044497 - 0.13696970 i$ &
$7.70288435 - 0.14015649 i$ &
$3.66653$ &
$2.32664$ \\
$0.15$ &
$8.55749294 - 0.16465262 i$ &
$9.67890740 - 0.17769503 i$ &
$13.1045$ &
$7.92117$ \\
$0.20$ &
$9.59681747 - 0.19152351 i$ &
$12.58708647 - 0.22615793 i$ &
$31.159$ &
$18.0836$ \\\hline
\multicolumn{5}{ |c| }{$m M= 0.30$} \\ \hline
$M\beta$ & $\text{Pseudospectral\, method} $ & $ \text{WKB}$ & $ \epsilon_{Re(\omega)} $ & $ \epsilon_{Im(\omega)} $ \\\hline
$-0.06$ &
$2.08425622 - 0.03054706 i$ &
$2.25345872 - 0.03266898 i$ &
$8.11812$ &
$6.9464$ \\
$-0.01$ &
$4.35735346 - 0.07029005 i$ &
$4.35741099 - 0.07029073 i$ &
$0.0013203$ &
$0.00096742$ \\
$0.01$ &
$5.01923188 - 0.08343171 i$ &
$5.01927986 - 0.08343226 i$ &
$0.000955923$ &
$0.000659222$ \\
$0.10$ &
$7.43233530 - 0.13691473 i$ &
$7.70492677 - 0.14010254 i$ &
$3.66764$ &
$2.32832$ \\
$0.15$ &
$8.55946661 - 0.16458739 i$ &
$9.68150472 - 0.17763367 i$ &
$13.1087$ &
$7.92666$ \\
$0.20$ &
$9.59884736 - 0.19144971 i$ &
$12.59077885 - 0.22609358 i$ &
$31.1697$ &
$18.0955$  \\\hline
\end {tabular}
\end {table}

\newpage
\subsection{The de Sitter family}

The de Sitter family of modes, are related to the accelerated expansion of the Universe, which is related to the surface gravity of the cosmological horizon of pure dS space. They correspond to purely imaginary modes which can be very well approximated by the pure dS QNMs. The pure dS modes are determined by two branches \cite{Du:2004jt, LopezOrtega:2006my}, and the most important for our purposes corresponds to the lowest lying solution, given by
\begin{align}
\label{dS1}
\omega_{\text{pure dS}} =-i \sqrt{\frac{\Lambda}{3}} (2n+ \ell+ \frac{3}{2} \pm \sqrt{\frac{9}{4}- 3 \frac{m^2}{\Lambda}})~,
\end{align}
where $n$ is the overtone number.

\begin{figure}[h!]
\begin{center}
\includegraphics[width=0.55\textwidth]{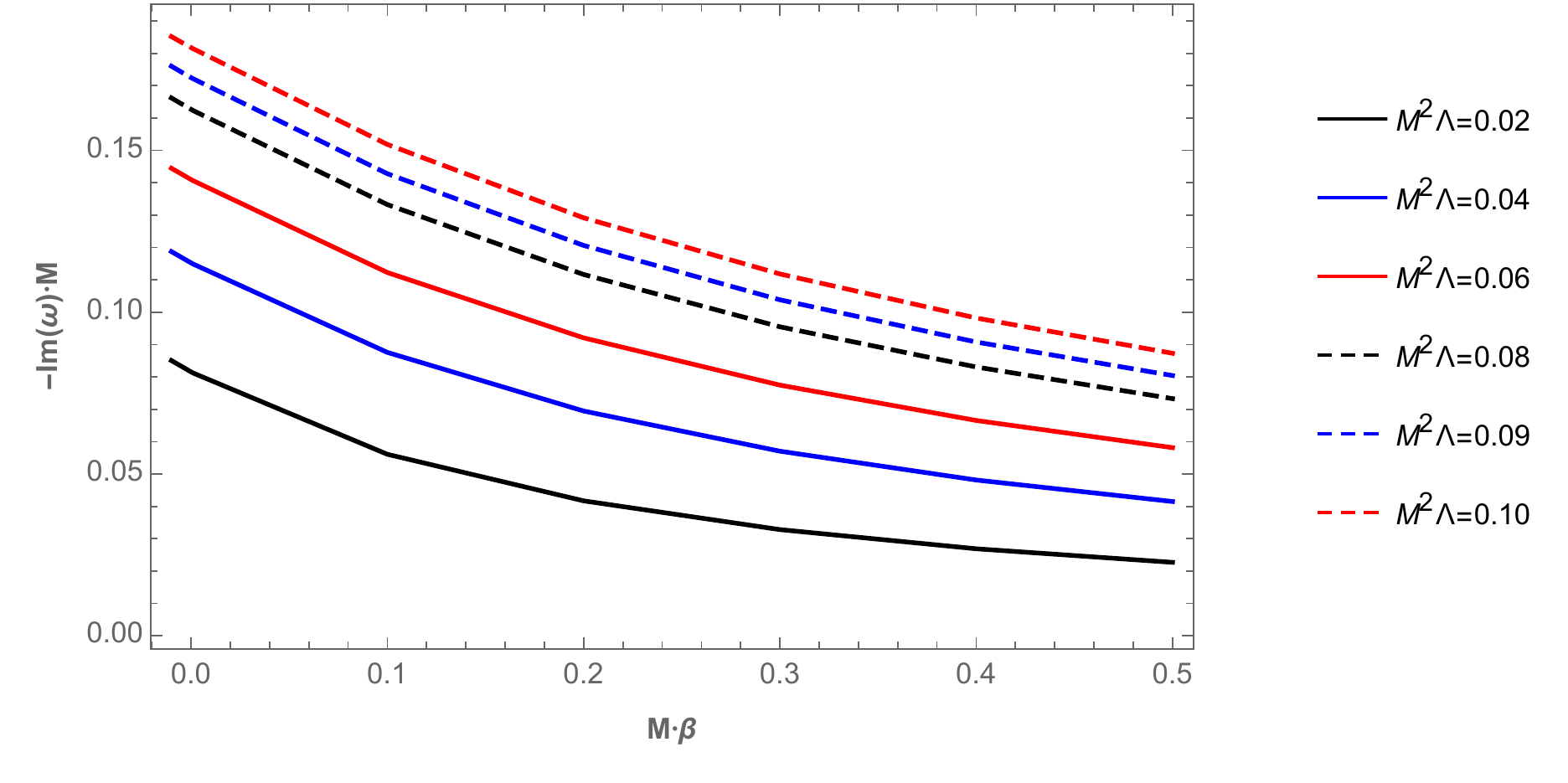}
\end{center}
\caption{The behavior of the fundamental dS mode ($n_{dS}=0$)
$-Im(\omega) M$ as a function of $M \beta$ obtained by using the pseudospectral Chebyshev method for massless scalar field with $\ell=1$ and different values of the cosmological constant.}
\label{F7}
\end{figure}

In order to visualize the behavior of the QNFs for the dS modes, in Fig. \ref{F7} we plot $-Im(\omega) M$ for $\ell=1$ and different values of the cosmological constant, note that there is a slowly decay rate when $M \beta$ increases, and when the cosmological constant decreases.
Also, as before, we will study the behavior of the dS QNFs as a function of the scalar field mass. So, we obtain some QNFs for different values of the parameters $\ell$ and $n_{dS}=0$ \footnote{$n_{dS}$ corresponds to  the overtone number of de Sitter modes}, for a fixed values of the black hole mass and the cosmological constant, see table \ref{dS1} for $M \beta=0.01$. As we mentioned, this family is dominant for small masses and  it
can acquires a real part becoming complex, which depend on the scalar field mass and the angular number. Also, it is possible to observe in the table \ref{dS1} that the decay rate always increases when the angular number increases, thereby the longest-lived modes are the ones with
lower angular number and the anomalous behavior is not present in this family.

\begin {table}[ht]
\caption {Fundamental dS modes ($n_{dS} =
      0 $) for massive scalar fields with $\ell = 0, 1, 2 $  in the background of black hole solution of $f(R)$ gravity with $M^2 \Lambda = 0.04 $, and $M \beta=0.01$.}
\label {dS1}\centering
\begin {tabular} { | c | c | c | c |}
\hline
$\ell$ & $m M = 0.02 $ & $m M= 0.05 $ & $m M= 0.10 $ \\\hline
$0$ &
$-0.00096728 i$ &
$-0.00623134 i$ &
$-0.02841651 i$  \\
$1$ &
$-0.11303750 i$ &
$-0.11914334 i$ &
$-0.14367078 i$ \\
$2$ &
$-0.22668753 i$ &
$-0.23290179 i$ &
$0.39967626 - 0.25614428 i$  \\\hline
$\ell$ & $m M= 0.15 $ & $m M= 0.20 $ & $m M= 0.25 $ \\\hline
$0$ &
$ -0.09297689 i$ &
$0.09799882 - 0.17514088 i$ &
$0.14338038 - 0.17214279 i$  \\
$1$ &
$-0.20354608 i$ &
$0.23373131 - 0.25290461 i$ &
$0.22754315 - 0.23202223 i$  \\
$2$ &
$0.40221808 - 0.25380624 i$ &
$0.40581565 - 0.25048547 i$ &
$0.41051976 - 0.24611626 i$ \\\hline
$\ell$ & $m M= 0.30 $ & $m M= 0.35 $ & $m M= 0.40 $ \\\hline
$0$ &
$0.17714386 - 0.16988129 i$ &
$0.20756224 - 0.16809914 i$ &
$0.23661390 - 0.16676332 i$  \\
$1$ &
$0.25091348 - 0.20524307 i$ &
$0.27737304 - 0.19237303 i$ &
$0.30313974 - 0.18446717 i$  \\
$2$ &
$0.41643654 - 0.24058189 i$ &
$0.42382467 - 0.23371594 i$ &
$0.43324930 - 0.22546868 i$  \\\hline
\end {tabular}
\end {table}

\subsection{The dominance family}

\begin{figure}[h]
\begin{center}
\includegraphics[width=0.6\textwidth]{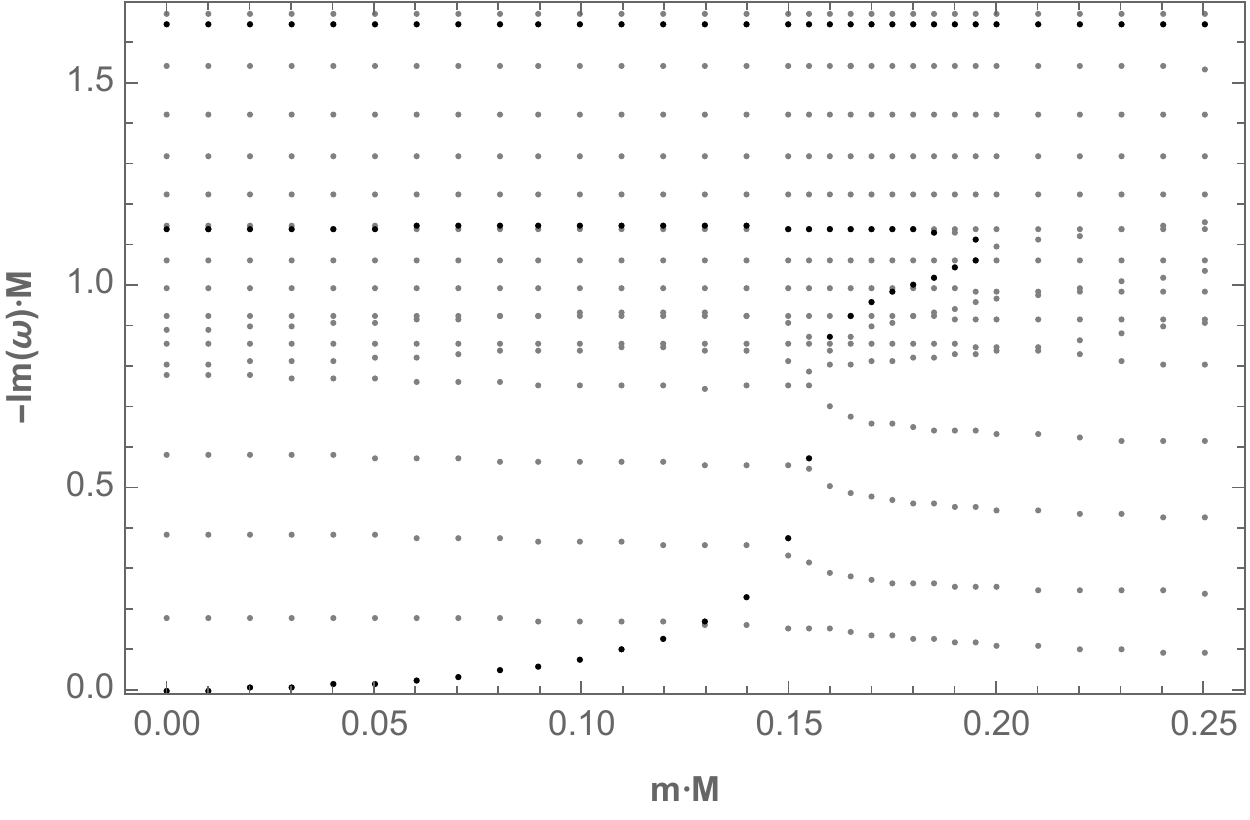}
\end{center}
\caption{The behavior of the imaginary part of the quasinormal frequencies  $-Im(\omega) M$ as a function of the scalar field mass $m M$ for different overtone numbers with $\ell=0$, $M \beta=0.01$, and $M^2 \Lambda=0.04$. Black points for purely imaginary QNFs and gray  points for complex QNFs.}
\label{FP1}
\end{figure}

In order to observe the dominance between the photon and de Sitter family
we plot in Fig. \ref{FP1}
the imaginary part of the QNFs as a function of the scalar field mass, for different overtone numbers and $\ell=0$, where the black points correspond to the purely imaginary QNFs, and the gray color points correspond to the complex QNFs for small values of the parameter $M \beta=0.01$. The behavior is similar to the observed in Ref. \cite{Aragon:2020tvq}.
We can recognize the two families for zero mass of the scalar field, a family of complex QNFs, and a purely imaginary family. As we have seen, the purely imaginary modes belong to the family of de Sitter modes,
while the complex ones
corresponds to the photon sphere modes.

In order to interpret this figure, first observe the black points, and $n_{dS}=0$.
This family
is dominant for small masses and near $m M=0.17$ it
can acquires a real part becoming complex due to the continuity between the black and gray points, and also
could be connect with the dS modes
with $-Im(\omega)M\approx 1$. Also, for low values of the mass, it is possible to observe that for $n_{dS}=1$ and $n_{dS}=2$,
can combine acquiring a real part becoming complex for $m M \approx 0.05$; the same occurs for the overtone numbers $n_{dS}=3$ and $n_{dS}=4$, and for $m M \approx 0.025$  becoming complex. Besides, for higher overtones numbers, there are dS modes purely imaginary
for all the range of mass considered. On the other hand, for the photon sphere modes, gray points, we can see that for $n_{PS}=0$ and small mass this branch is subdominant; however, for $m M \approx 0.155$ it begins to dominate.

\begin{figure}[h]
\begin{center}
\includegraphics[width=0.6\textwidth]{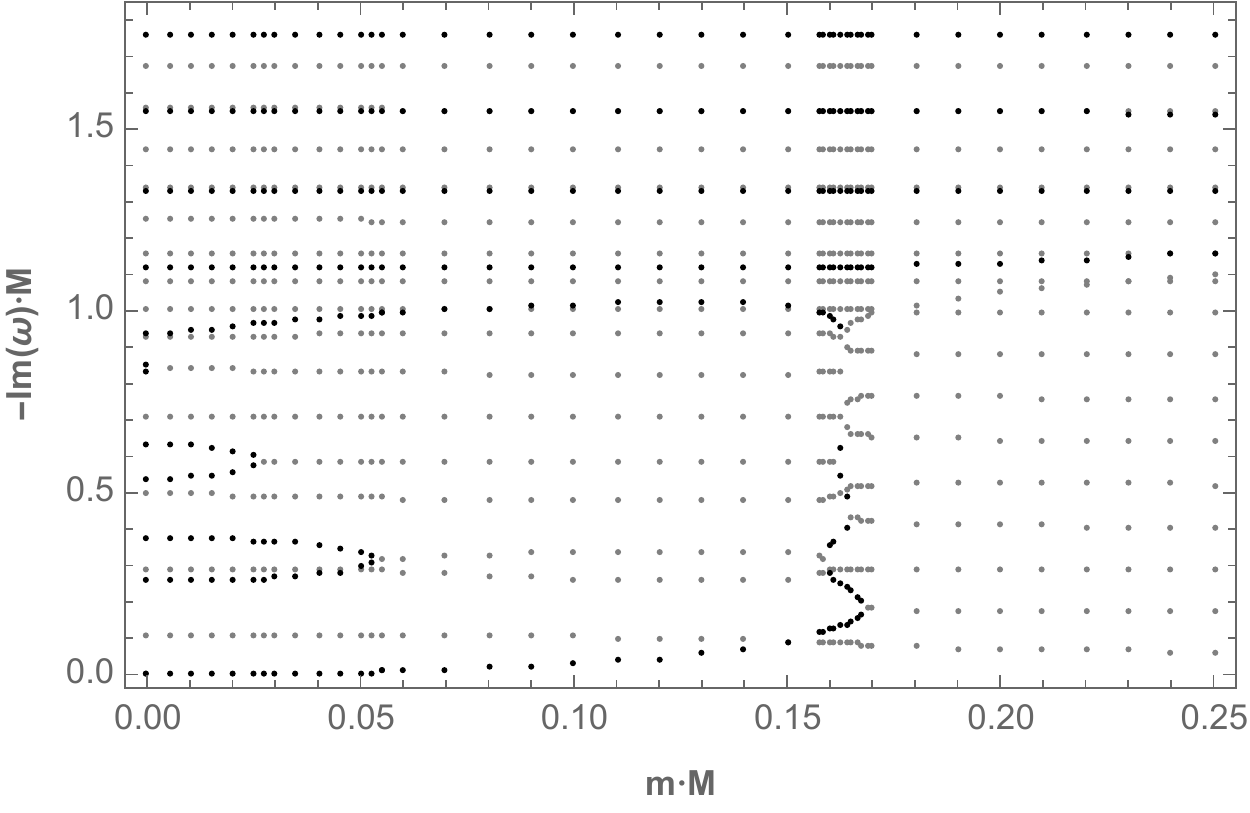}
\end{center}
\caption{The behavior of the imaginary part of the quasinormal frequencies $-Im(\omega) M$ as a function of the scalar field mass $m M$ for different overtone numbers with $\ell=0$, $M \beta=0.20$, and $M^2 \Lambda=0.04$. Black points for purely imaginary QNFs and gray points for complex QNFs.}
\label{FP2}
\end{figure}

Now, for a bigger value of the parameter $M \beta=0.20$, see Fig. \ref{FP2}.
As before, the black points correspond to the purely imaginary QNFs, and the gray points correspond to the complex QNFs. Note that for low values of the scalar field mass dS modes
are dominant; however for larger mass the photon sphere modes
are dominant. Also, it is possible to observe that
dS modes with $n_{dS}=0$
could to connect with dS modes
with $n_{dS}=1$ (
$-Im(\omega) M \approx 1.10$). Besides, for higher overtones numbers, there is a purely imaginary family of dS modes for all the range of mass considered. Also, for the photon sphere modes, colored points, we can see that for $n_{PS}=0$ and small mass this family is subdominant; however, for $m M  \approx 0.13$ it begins to dominate.

\newpage
\section{Conclusions}
\label{conclusion}

In this work, we considered an asymptotically de-Sitter black hole solution of a specific $f(R)$ modified gravity as background. In order to see if strong curvature effects, that arise from non-linear terms in $R$ in the action, can modify the behaviour of the QNMs. Thus, we studied the propagation of a probe scalar field and we analyzed the presence of anomalous behaviour in the quasinormal modes spectrum. First, we showed that the QNMs spectrum consist of two families: the photon sphere modes and the de Sitter modes, and their behaviour
depends on the parameter $\beta$ which characterizes the considered black hole solution of the $f(R)$ gravity.

For the photon sphere modes we found that for massless scalar field and small values of the parameter $M \beta$, the decay rate decreases when the cosmological constant increases, the behaviour is similar to the observed for Schwarzschild-dS black holes \cite{Aragon:2020tvq}. On the other hand, for large values of the parameter $M \beta$ the behaviour is inverted, i.e. the decay rate increases when the cosmological constant increases.

 Also, we found that the photon sphere modes show an anomalous behaviour that depends on the value of the parameter $\beta$ and the mass of the scalar field. As in the case of the Schwarzschild-dS black holes, which corresponds to $\beta=0$, there is a critical value of the mass of the scalar field, which depends on the cosmological constant, beyond which the QNMs have an anomalous behaviour. However, in the $f(R)$ gravity this critical value depends, not only on cosmological constant but also on the new parameter $\beta$, which expresses the presence of non-linear terms in the curvature. We found that the critical value of the scalar mass decreases when the parameter $M \beta$ increases, and there is a critical value of $M \beta$ where the critical mass is zero indicating that
 there is not an anomalous behaviour of the QNMs.
  Also, we saw that the critical value of the mass of the scalar field  increases when the overtone number increases; however when the parameter $M \beta$ approaches a near extremal black hole, i.e. when the cosmological horizon approximates to the event horizon, the critical value of the mass  of the  scalar field does not depend of the overtone number.

On the other hand, for the dS modes we observed that for massless scalar field they are purely imaginary, and for massive scalar field they can acquire an oscillatory part becoming complex for some value of the mass and higher. Also, we do not found an anomalous behaviour for this family of modes.

Also, we found that the imaginary part of the quasinormal frequencies is always negative leading to a stable  propagation of the scalar fields in this background for both families of modes. The PS family or the dS family may dominates the evolution of the perturbation depending on the values of the parameters. We showed that for small values of $M^2 \Lambda$ the dS modes
dominate for small values of the scalar field mass, while that the photon sphere modes
dominate for larger values of the scalar field mass.
Also, the decay rate of the PS modes increase with $M \beta$, while that the decay rate of the dS modes decrease, so it can occurs that for low values of $M \beta$ the PS modes dominate and for high values of $M \beta$ the dS modes dominate, depending on the value of $M^2 \Lambda$.

An interesting extension of this work is to calculate the QNMs and QNFs of the $f(R)$ gravity theory if we allow the matter fields to backreact with the metric. Then we can study the properties of the resulting hairy black holes \cite{Karakasis:2021lnq, Tang:2020sjs} and possibly see the interplay effects of the geometric scalar present in the $f(R)$ theory and the matter scalar field minimally coupled to the modified gravity theory.

\acknowledgments
P. A. G. acknowledges the hospitality of the Universidad de La Serena where part of this work was undertaken. Y.V. acknowledge support by the Direcci\'on de Investigaci\'on y Desarrollo de la Universidad de La Serena, Grant No. PR18142.

\end{document}